\documentclass{aastex63}

\begin{document}

\title{Nuclear Ignition of White Dwarf Stars by Relativistic Encounters with Rotating Intermediate Mass Black Holes}

\author{Peter Anninos}
\affil{Lawrence Livermore National Laboratory, Livermore, CA 94550, USA}
\author{Robert D. Hoffman}
\affil{Lawrence Livermore National Laboratory, Livermore, CA 94550, USA}
\author{Manvir Grewal}
\affil{Department of Physics, Columbia University, New York, NY 10027, USA}
\author{Michael J. Lavell}
\affil{Lawrence Livermore National Laboratory, Livermore, CA 94550, USA}
\author{P. Chris Fragile}
\affil{Department of Physics and Astronomy, College of Charleston, Charleston, SC 29424, USA }
\affil{Kavli Institute of Theoretical Physics, University of California at Santa Barbara, Santa Barbara, CA 93106, USA}

\begin{abstract}
We present results from general relativistic calculations of nuclear ignition in white dwarf
stars triggered by near encounters with rotating intermediate mass black holes
with different spin and alignment parameters. These encounters create thermonuclear environments
characteristic of Type Ia supernovae capable of producing both calcium and iron group elements in arbitrary ratios,
depending primarily on the proximity of the interaction which acts as a strong moderator of nucleosynthesis.
We explore the effects of black hole spin and spin-orbital alignment on burn product synthesis
to determine whether they might also be capable of moderating reactive flows.
When normalized to equivalent impact penetration, accounting for frame dragging corrections,
the influence of spin is weak, no more than 25\% as measured by nuclear
energy release and mass of burn products, even for near maximally rotating black holes.
Stars on prograde trajectories approach closer to the black hole 
and produce significantly more unbound debris and iron group elements 
than is possible by encounters with nonrotating black holes or by retrograde orbits, 
at more than 50\% mass conversion efficiency. The debris contains
several radioisotopes, most notably $^{56}$Ni,
made in amounts that produce sub-luminous (but still observable) light curves compared to branch-normal SNe Ia.
\end{abstract}

\keywords{Black holes --- White dwarf stars --- Black hole physics --- Hydrodynamics --- Explosive nucleosynthesis}

\section{Introduction}

The tidal disruption (TD) of white dwarf (WD) stars by intermediate mass black holes (IMBH) are
violent cosmic events capable of generating observable electromagnetic and gravitational wave energies
\citep{Rees88, Haas12, Macleod16}. One particularly intriguing aspect of TD events is the possibility
that compressive forces exerted by the BH might trigger explosive thermonuclear reactions inside the WD, producing
isotopically rich environments composed of both intermediate and iron-group nuclei \citep{Luminet89,Rosswog09}. 
If sufficient amounts of radioactive nuclei
are synthesized and dispersed among unbound debris, their decay and subsequent reprocessing
could give rise to observable transient emissions.
And if these signatures are clearly distinguishable from other known transient patterns (e.g., Type Ia supernovae) they
will provide strong evidence to the existence of IMBHs and help determine their mass function
\citep{Gerssen02, Gerssen03, Gebhardt02, Gebhardt05, Dong07}.

A host of simulated WD-IMBH binary systems have demonstrated that
indeed explosive nucleosynthesis is a likely outcome for events with sufficiently 
large tidal strength, defined as $\beta=R_T/R_P$, the ratio of tidal to perihelion radii
\citep{Rosswog09, Tanikawa17, Kawana18, Anninos18}. The precise ignition threshold depends
on the WD mass, but a general requirement is $\beta >3$ (5) for the more (less) massive WD stars.
These scenarios additionally have the potential to release more nuclear 
energy than the star's quiescent binding energy and give rise to short-burst
accretion rates of up to $10^7$ - $10^8 M_\odot$ yr$^{-1}$ \citep{Haas12,Anninos18}.
For weaker or greater perihelion radius interactions, nucleosynthesis might still occur but not as efficiently.
The outcome in these cases will likely produce calcium-rich (as opposed to iron-rich) debris
depending on the WD mass and interaction strength \citep{Holcomb13,Sell15,Anninos18}.
These conclusions are gathered from a collective body of calculations that considered somewhat
idealized WD-IMBH configurations and (for the most part) pseudo-Newtonian interactions. 
The effects of BH spin and spin-axis orientation on the thermonuclear environment of the WD,
for example, have not been investigated. Black hole spin was considered by \citet{Haas12} and \citet{Evans15}
with general relativity, and by \citet{Gafton19} with a generalized Newtonian
method, but none of those studies included reactive networks in their calculations.

Here we study nuclear processes triggered by the disruption of WD stars by close encounters with rotating and tilted IMBHs, i.e. 
BHs with spin axes not aligned perpendicular to the orbital plane. 
Our goal is to extend previous work by considering a broader range of interaction scenarios,
and account for general relativistic effects expected from near encounters.
To this end, we use a combined moving mesh and adaptive mesh refinement (AMR)
capability that we developed specifically for this problem class \citep{Anninos12,Anninos18}.
The AMR/moving mesh hybrid approach allows us to move the base grid along Lagrangian fluid lines,
simultaneously following the Keplerian trajectory and compression of the WD stars while at the same time refining
on the densest parts of the stars as they compress and stretch from tidal forces. The enhanced grid
resolution made possible by this hybrid adaptive approach is critical in order to
achieve sufficient zone resolution along the orbital plane and provide the necessary scale height coverage to
resolve reactive flows and internal shock features \citep{Tanikawa17}. Our numerical methods and calculations
are implemented and carried out in a background metric approximation.
We solve the general relativistic hydrodynamics equations in a background Kerr-Schild
spacetime written in Cartesian coordinates, modified to account for the evolving gravitational potential of the white dwarf, and
coupled with an $\alpha$-chain nuclear network and burn energy coupling.

We begin in Section \ref{sec:methods} with a brief discussion of our numerical methods, physical models
(equation of state, reactive networks, initial data, etc.), and dynamic mesh strategy tuned to approach the high
spatial resolution required for resolving nucleosynthesis.
Our results follow in Section \ref{sec:results}, and we conclude with a brief summary
in Section \ref{sec:conclusions}.

\section{Methods and Models}
\label{sec:methods}

All calculations are performed with the {\sc Cosmos++} code
\citep{Anninos05,Fragile12,Fragile14,Anninos17}, which solves
the equations of general relativistic hydrodynamics coupled with
thermonuclear reactions and energy generation on unstructured, moving and adaptively refined (AMR) meshes.  
Except for one addition to be described immediately below, the combined AMR and 
grid motion strategy is for the most part identical to what we presented in
\citet{Anninos12,Anninos18}, so we do not repeat those details here. We do however remind
the reader that our highest resolution calculations use a base grid of $96^3$ cells and two additional
refinement levels for an effective $384^3$ resolution of $\sim10^7$ cm along the $x$ and $y$ coordinates
within the orbital plane. Our vertical mesh motion strategy additionally follows
the Lagrangian compression down to a scale of $\lesssim10^6$ cm (or equivalently
1/300th of a Schwarzschild radius) along the $z$ axis perpendicular to the orbital plane.

The domain of the grid is initially constructed to cover 10 (5) stellar radii parallel
(perpendicular) to the orbital plane, but we emphasize that the domain extent varies
in both time and space as the mesh velocity adjusts anisotropically to follow tidal debris.
Boundary conditions for the hydrodynamics on the outer edges of the grid are set to outflow:
all ghost zone quantities are equated to the values of their adjacent internal-zone neighbors,
except the velocity component normal to the boundary is set to zero if it points onto the grid. 
We use multipole boundary conditions for the gravitational potential, up to and including 10 moments.
The time step is limited by the minimum stability criteria for the hydrodynamics, dynamical self-gravity,
nuclear energy production, and fractional abundance changes \citep{Anninos18}.

In order to accommodate the possibility that an arbitrarily aligned BH spin axis might entrain material
off the orbital plane we have generalized the vertical grid motion to expand the mesh if
the stellar boundary (defined as $10^{-3}$ of the peak density) is detected near the outer grid boundaries.
This is accomplished by setting the vertical mesh velocity component to
\begin{equation}
|V_g^z(t)| = \chi_0 ~\left| V^z_{L}(t)\right| ~\left|\frac{z(t)}{L_z(t)} \right|^{\chi_N} ~,
\label{eqn:meshvel}
\end{equation}
where $z(t)$ is the time-dependent coordinate,
$L_z(t)$ is the changing vertical length of the grid along the positive $z$-axis
(distance from the orbital plane to the outer boundary),
$\chi_0 \ge 0$ is a constant multiplier, and
$|V^z_{L}(t)|$ is the magnitude of the $z$-component of the star's maximum 
outward-directed Lagrangian velocity.
The factor $|z(t)/L_z(t)|^{\chi_N}$ smoothly scales back the vertical grid motion
to zero at the midplane, preventing loss of resolution in and around the orbital plane.
Values of the exponent $\chi_N$ greater (less) than unity tend to move the outer grid edges
faster (slower) than the midplane cells. We find $\chi_N=1.2$ and $\chi_0 = 0.25$ work reasonably well.
This grid velocity component is applied symmetrically above and below the midplane, but
only for tilted spin cases and only when significantly overdense material reaches the boundaries.
Equation(\ref{eqn:meshvel}) is similar in form to what we use for collapsing the mesh
onto the orbital plane at early times, but we now also adapt it (with different parameter sets)
to expand the mesh and track off-plane entrainment.

In addition to generalizing the mesh motion strategy to accommodate off-plane entrainment,
we have made a number of other significant improvements to the physical models.
The most important of which is a more accurate equation of state (EOS),
upgraded from an idealized 2-state (hot, cold) and 2-component (ion, radiation) treatment
to a Helmholtz model suited for representing
electron degeneracy and relativistic and electron-positron contributions.
Our implementation is based on the Torch code \citep{Timmes99,Timmes00a} and is
designed to work with arbitrary isotopic compositions and can be coupled
to nuclear reaction networks as we have done for this work. The Helmholtz EOS is utilized
in tabular form with densities spanning the range $10^{-12} \le \rho \le 10^{15}$ gm cm$^{-3}$,
and temperatures $10^{3} \le T \le 10^{13}$ K. Interpolations between table entries are performed with
biquintic Hermite polynomials that ensures thermodynamic consistency.

Another improvement concerns the development of a particle tracer capability that we use to sample and
track stellar material and its thermodynamic state (density, temperature, energy) as it evolves in space and time.
This data (consisting of tens of thousands of particles) 
is stored at regular cyclic intervals and, upon completion of the calculations,
is input to a stand-alone network solver code for post-processing with a much larger and more accurate nuclear
reaction network than is possible inline. For the inline network we use the same 19-isotope $\alpha$-chain
and heavy-ion reaction model that we used previously \citep{Weaver78,Timmes99,Timmes00b,Anninos18}.
It is fully coupled with the relativistic hydrodynamics equations, including isotopics and nuclear energy released.
The 19-isotope network was the main diagnostic we relied on for calculating burn products
in our previous work. We continue to rely on this model for nuclear energy feedback
to the hydrodynamics, but now we use the off-line post-processing
capability to calculate more detailed isotopic compositions using a version of the Torch code
\citep{Timmes99} that we modified for parallel processing.

We have also accounted for self-gravity beyond the quasi-static Cowling approximation 
used in our previous work.
This is done by solving for the gravitational potential ($\phi$) of the WD 
in an asymptotically flat spacetime at each time cycle, then
adding it as a first order perturbation to the Cartesian Kerr-Schild metric, which we 
write in {\it untilted} form assuming unit light speed
\begin{equation}
ds^2_{\mathrm{(KS)}} = (\eta_{\mu\nu} + f p_\mu p_\nu) dx^\mu dx^\nu =
      -\left(1 - f \right) dt^2 + 
       \left(f p_i\right) dt dx^i +
       \left(\delta_{ij} + f p_i p_j\right) dx^i dx^j ~,
\end{equation}
where $f = 2GM_{\mathrm{BH}} \tilde{r}^3/(\tilde{r}^4 + a^2 \tilde{z}^2)$, 
$M_{\mathrm{BH}}$ is the BH mass, $a$ is its spin,
$p_i$ are the spin modified directional cosines
\begin{eqnarray}
p_0 &=& 1  ~, \nonumber \\
p_x &=& (\tilde{r} \tilde{x} + a \tilde{y})/(\tilde{r}^2 + a^2) ~, \nonumber \\
p_y &=& (\tilde{r} \tilde{y} - a \tilde{x})/(\tilde{r}^2 + a^2) ~, \nonumber \\
p_z &=& \tilde{z}/\tilde{r} 	~,
\end{eqnarray}
$\tilde{x}^i$ define the tilted coordinates
\begin{eqnarray}
\tilde{x}     &=&  x \cos(\zeta) + z \sin(\zeta) ~, \nonumber \\
\tilde{y}     &=&  y ~, \nonumber \\
\tilde{z}     &=& -x \sin(\zeta) + z \cos(\zeta) ~, 
\end{eqnarray}
and
\begin{eqnarray}
\tilde{r}_*^2 &=& \tilde{x}^2 + \tilde{y}^2 + \tilde{z}^2 - a^2  ~, \nonumber \\
\tilde{r}^2   &=& \frac{1}{2} ( \tilde{r}_*^2 + \sqrt{\tilde{r}_*^4 + 4 a^2 \tilde{z}^2} ~)  ~.
\end{eqnarray}
$\zeta$ is the angle of tilt that we have arbitrarily assigned around the $y$ axis.
The complete {\it tilted} spacetime metric is written in terms of the {\it untilted} metric as
\begin{eqnarray}
g_{00}^{(\mathrm{TKS})} &=& g_{00}^{(\mathrm{KS})} ~,	\nonumber \\
g_{02}^{(\mathrm{TKS})} &=& g_{02}^{(\mathrm{KS})} ~,	\nonumber \\
g_{22}^{(\mathrm{TKS})} &=& g_{22}^{(\mathrm{KS})} ~,	\nonumber \\
g_{01}^{(\mathrm{TKS})} &=& \cos(\zeta) g_{01}^{(\mathrm{KS})} - \sin(\zeta) g_{03}^{(\mathrm{KS})} ~,	\nonumber \\
g_{12}^{(\mathrm{TKS})} &=& \cos(\zeta) g_{12}^{(\mathrm{KS})} - \sin(\zeta) g_{23}^{(\mathrm{KS})} ~,	\nonumber \\
g_{03}^{(\mathrm{TKS})} &=& \sin(\zeta) g_{01}^{(\mathrm{KS})} + \cos(\zeta) g_{03}^{(\mathrm{KS})} ~,	\nonumber \\
g_{23}^{(\mathrm{TKS})} &=& \sin(\zeta) g_{12}^{(\mathrm{KS})} + \cos(\zeta) g_{23}^{(\mathrm{KS})} ~,	\nonumber \\
g_{11}^{(\mathrm{TKS})} &=& \cos(\zeta)\cos(\zeta) g_{11}^{(\mathrm{KS})} + \sin(\zeta)\sin(\zeta) g_{33}^{(\mathrm{KS})}
                 -(\sin(\zeta)\cos(\zeta) + \sin(\zeta)\cos(\zeta)) g_{13}^{(\mathrm{KS})} ~,	\nonumber \\
g_{13}^{(\mathrm{TKS})} &=& \cos(\zeta)\sin(\zeta) g_{11}^{(\mathrm{KS})} - \cos(\zeta)\sin(\zeta) g_{33}^{(\mathrm{KS})}
                 +(\cos(\zeta)\cos(\zeta) - \sin(\zeta)\sin(\zeta)) g_{13}^{(\mathrm{KS})} ~,	\nonumber \\
g_{33}^{(\mathrm{TKS})} &=& \sin(\zeta)\sin(\zeta) g_{11}^{(\mathrm{KS})} + \cos(\zeta)\cos(\zeta) g_{33}^{(\mathrm{KS})}
                 +(\sin(\zeta)\cos(\zeta) + \sin(\zeta)\cos(\zeta)) g_{13}^{(\mathrm{KS})}	~.
\end{eqnarray}
Neglecting high order corrections that come from spin and near-field effects, 
the WD potential contributes to the scalar curvature components of the metric in the manner
$g_{\mu\nu} = \eta_{\mu\nu} + f p_\mu p_\nu -2\phi \delta_{\mu\nu}$, so that the
line element becomes finally
\begin{equation}
ds^2 = \left(g_{00}^{(\mathrm{TKS})} - 2\phi \right) dt^2 + 
       \left(g_{0i}^{(\mathrm{TKS})} \right) dt dx^i +
       \left(g_{ij}^{(\mathrm{TKS})} - 2 \phi \delta_{ij} \right) dx^i dx^j ~.
\end{equation}
We find this to be a reasonable approximation that maintains equilibrium 
outside of the tidal radius. Of course it is increasingly less
accurate near the BH, but there it is also less important. We apply the
perturbation potential early on in an encounter until the WD approaches to about a half tidal radius of
the BH, at which point self-gravity can be safely ignored to speed up the calculations.

This treatment of the spacetime metric provides flexibility for the alignment of the BH spin axis
while confining the WD trajectory to the $z=0$ plane.
Alignment of the WD trajectory to one of the principle axes of the grid mitigates some of the effects
of numerical dissipation and mesh imprinting on its evolution that would be more strongly present
if the WD were to travel along the cell diagonals, which would be the case if we had instead chosen
to align the BH spin axis to the grid. It also simplifies the grid motion strategy for collapsing
zones along Lagrangian fluid lines orthogonal to the orbital plane.

\subsection{Initial Data}
\label{subsec:initialData}

In \citet{Anninos18}, henceforth referred to as Paper 1, we considered 0.2 and 0.6 solar mass WD stars approaching
nonrotating intermediate mass black holes, varying the perihelion distance to achieve tidal strengths
between $2.6 \leq \beta \leq 17$. Here we extend the interaction
parameter space by considering black holes with different spins and rotation axis orientations
(tilt) to see if correlations exist between these parameters and nuclear activity.
In order to reduce the number of calculations to a manageable level, we constrain the parameter space
to two basic scenarios: a $0.6 M_\odot$ (CO) WD approaching a $10^3 M_\odot$ BH (case A), and a
$0.15 M_\odot$ (He) WD approaching a $10^4 M_\odot$ BH (case B). Black hole spin is additionally limited to
either zero or 0.9 (in BH mass units, where 1 is a maximally rotating Kerr BH), 
and three tilt configurations corresponding to retrograde, polar, and prograde orbital alignments.

Although our maximum spin case has an $a/M$ that is less than the theoretical maximum of 0.998 \citep{Thorne74}, a 
value close to 0.9 is more consistent with estimates of realistic spin equilibrium values that account for 
spin-up and spin-down processes \citep{Gammie04}. Furthermore, techniques for measuring black hole 
spin have yet to converge, with the Fe-line fitting method tending to favor high spins 
($a/M \gtrsim 0.8$) \citep{Reynolds14}, the thermal spectral fitting method favoring more modest 
spins ($0.2 \lesssim a/M \lesssim 0.9$) \citep{McClintock14}, and early gravitational wave detections possibly 
suggesting very low spins ($|a/M| \le 0.3$) \citep{Farr17}. Given this uncertainty, $a/M = 0.9$ seems like a 
safe, and reasonable, value to consider.

We also limit our studies to a narrow range of interaction strengths $4.4 \le \beta \le 6.8$,
where $\beta$ is the ratio of tidal to perihelion radii $\beta=R_T/R_P$, with tidal radius
\begin{eqnarray}
R_\mathrm{T} &\approx& 1.2\times10^{5} \left(\frac{R_\mathrm{WD}}{10^9 \mathrm{cm}}\right)
                       \left(\frac{M_\mathrm{BH}}{10^3 M_\odot}\right)^{1/3}
                       \left(\frac{M_\mathrm{WD}}{0.6 M_\odot}\right)^{-1/3}	~\mathrm{km} ~,	\\
             &\approx& 0.8\times10^{2} \left(\frac{R_\mathrm{WD}}{10^9 \mathrm{cm}}\right)
                       \left(\frac{M_\mathrm{BH}}{10^3 M_\odot}\right)^{-2/3}
                       \left(\frac{M_\mathrm{WD}}{0.6 M_\odot}\right)^{-1/3}	~R_G ~,
\end{eqnarray}
for a black hole mass $M_\mathrm{BH}$, stellar radius $R_\mathrm{WD}$, and stellar mass $M_\mathrm{WD}$.
$R_G$ is a convenient scale length defined by the gravitational radius
\begin{equation}
R_G = \frac{GM_\mathrm{BH}}{c^2} \approx 1.5\times10^3 \frac{M_\mathrm{BH}}{10^3 M_\odot} ~\text{km}  ~.
\end{equation}

Table \ref{tab:runs} shows the different case studies considered in this report, focusing
on five principle parameters: BH mass ($M_\odot$), WD mass ($M_\odot$), 
black hole spin (BH mass units), spin-orbit alignment, and interaction strength ($\beta$).
Also shown are the star radii $R_{\mathrm{WD}}$,
the actual calculated (relativistically corrected) perihelion distance $R^*_P$,
and the relativistically corrected tidal strength $\beta^*$ experienced by the mass centroid of the WD in the simulations.
The run labels represent physical parameters: The first (letter) index signifies the [BH, WD] mass combination
(A$\equiv[10^3, 0.6]~M_\odot$, B$\equiv[10^4, 0.15]~M_\odot$), the second (number) index is the BH spin, and the third 
(letter) index identifies the orbit as retrograde, prograde, or polar (equivalently mid-plane alignment).
The index following the hyphenation is simply a running letter index representing different perihelion radii
and therefore different $\beta$.
The order in which the cases are presented in Table \ref{tab:runs} reflects the results
shown later in Table \ref{tab:results}: from highest peak temperature to lowest within each grouping.

The new physics elements incorporated into this work 
(Helmholtz EOS, particle tracers, self-gravity) come at significant computational
cost. In order to perform all the calculations needed to cover the parameter space we are forced
to carry out many of the calculations at less than ideal grid resolution, on a 96$^3$ grid
with two in place of three AMR levels. The high resolution calculations are run on a 96$^3$ grid
with three grid hierarchies and are identified with the same notation in Table  \ref{tab:runs} except they
are highlighted in bold text.
Although the grid resolution suffers by a factor of two
in directions parallel to the orbital plane, our grid motion strategy preserves
the vertical scale height resolution of $10^6$ cm in the orbital plane
at the expense of coarser zoning nearer grid boundaries.

\begin{deluxetable}{lccccccccc}
\tablecaption{Run Parameters \label{tab:runs}}
\tablewidth{0pt}
\tablehead{
\colhead{Run}             & 
\colhead{$M_\mathrm{BH}$} & 
\colhead{$M_\mathrm{WD}$} & 
\colhead{$R_\mathrm{WD}$} & 
\colhead{$R_\mathrm{P}$}  & 
\colhead{$R^*_\mathrm{P}$}& 
\colhead{Spin}            & 
\colhead{Orbit}           &
\colhead{$\beta$}         &
\colhead{$\beta^*$}       \\
                          & 
($M_\odot$)               & 
($M_\odot$)               & 
($R_G$)                   & 
($R_G$)                   & 
($R_G$)                   & 
($M_{BH}$)                & 
                          &
                          &
}
\startdata
\hline
A9p-c       & $10^3$ &  0.6  & 5.7  &  10    &   8.0   & 0.9  & prograde   & 6.8 & 8.5  \\
A9p-b       & $10^3$ &  0.6  & 5.7  &  12.5  &  10.7   & 0.9  & prograde   & 5.4 & 6.3  \\
A0-b        & $10^3$ &  0.6  & 5.7  &  13.5  &  10.6   & 0    & -                   & 5.0 & 6.4  \\
A9r         & $10^3$ &  0.6  & 5.7  &  15    &  10.8   & 0.9  & retrograde & 4.5 & 6.3  \\
A0-a        & $10^3$ &  0.6  & 5.7  &  15    &  12.3   & 0    & -                   & 4.5 & 5.5  \\
A9p-a       & $10^3$ &  0.6  & 5.7  &  15    &  13.3   & 0.9  & prograde   & 4.5 & 5.1  \\
A9m         & $10^3$ &  0.6  & 5.7  &  15    &  12.3   & 0.9  & polar      & 4.5 & 5.5  \\
\\
{\bf A9p-c} & $10^3$ &  0.6  & 5.7  &  10    &   8.1   & 0.9  & prograde   & 6.8 & 8.4  \\
{\bf A9r}   & $10^3$ &  0.6  & 5.7  &  15    &  10.7   & 0.9  & retrograde & 4.5 & 6.3  \\
{\bf A9p-a} & $10^3$ &  0.6  & 5.7  &  15    &  13.4   & 0.9  & prograde   & 4.5 & 5.1  \\
\\
B9p-b      & $10^4$ &  0.15 & 1.1  &   7    &   6.3   & 0.9  & prograde   & 6.3 & 7.0  \\
B0-b       & $10^4$ &  0.15 & 1.1  &   8    &   5.7   & 0    & -                   & 5.5 & 7.7  \\
B9r        & $10^4$ &  0.15 & 1.1  &  10    &   5.3   & 0.9  & retrograde & 4.4 & 8.3  \\
B0-a       & $10^4$ &  0.15 & 1.1  &  10    &   8.2   & 0    & -                   & 4.4 & 5.4  \\
B9m        & $10^4$ &  0.15 & 1.1  &  10    &   8.3   & 0.9  & polar      & 4.4 & 5.3  \\
B9p-a      & $10^4$ &  0.15 & 1.1  &  10    &   9.3   & 0.9  & prograde   & 4.4 & 4.7  \\
\enddata
\end{deluxetable}

Our choice of parameters is in part motivated by a compromise between proximity to
the black hole in order to maximize tidal and spin effects, and the scale height
resolution needed to adequately resolve nuclear processes in the orbital plane, a constraint
that depends strongly on the tidal strength $\delta z/R_{\mathrm{WD}} \sim \beta^{-3}$ \citep{Luminet86,Brassart08},
where $\delta z$ is the grid resolution perpendicular to the orbital plane.
This is the primary motivation for our selection of tidal strengths:
they are large enough to trigger robust nucleosynthesis but not
so large as to require excessive vertical resolution.

We point out that the ($\beta$, $R_P$) 
pairings quoted in Table \ref{tab:runs} differ slightly from Paper 1.
This is due to the different initial stellar models adopted for this work.
In another final departure from our previous work, we do not rely on solutions to the
relativistic Tolmon-Oppenheimer-Volkof (TOV) equations for the initial data.
Instead, we use hydrodynamic profiles generated from the MESA stellar evolution code \citep{Paxton11},
including density, internal energy, and isotopic compositions. These profiles do not differ
much (to within roughly a factor of two) from our older stellar models in density,
radial extent, and temperature. They do however differ enough that the same perihelion 
radius does not produce the same tidal strength characterization ($\beta$) presented in our previous studies.
Additionally the masses differ slightly as well. Although the 0.6 $M_\odot$ WD
is very close to what we considered previously (to within a few percent), the equivalent
$0.2 M_\odot$ model is now more precisely a 0.15 $M_\odot$ WD.
These are some of the reasons that we (re)consider nonrotating BH interactions
in this work (the upgrade to a Helmholtz EOS is another). 

As for isotopic compositions, the 0.6 $M_\odot$ model is composed of a central region with roughly a homogeneous
mixture of 1/3 $^{12}$C and 2/3 $^{16}$O with trace amounts of heavier nuclei, surrounded
in turn by layers of carbon-rich, helium-rich, and hydrogen-rich material, ordered from the inner core to 
the outer surface. The 0.15 $M_\odot$ model is composed of essentially a uniform distribution
of 99\% (by mass fraction) $^4$He with trace amounts of other species.
Considering the dominance of CO and He in the two stellar models, we occasionally refer the
0.6 (0.15) $M_\odot$ WD as CO (He) WD.
{\sc MESA} produces a much greater number of isotopes than appear in the inlined 19-isotope
burn model. The integrated mass fractions of those species is small (less than 1 to 2 percent,
depending on the He or CO WD)
so we ignore them and renormalize the sum of mass fractions to unity with a density weighted
scaling applied to each existing isotope.

\section{Results}
\label{sec:results}

The range of interaction strengths for case A encounters of a CO WD with a $10^3 M_\odot$ BH
are within the upper bound ($\beta_S = 8.4$) where the BH swallows the WD \citep{Kawana18}
\begin{equation}
\beta_S \lesssim 10 
                  \left(\frac{R_\mathrm{WD}}{10^9 \mathrm{cm}}\right)
                  \left(\frac{M_\mathrm{BH}}{10^3 M_\odot}\right)^{-2/3}
                  \left(\frac{M_\mathrm{WD}}{0.6 M_\odot}\right)^{-1/3}	~.
\end{equation}
Case B encounters, on the other hand, lie at or outside the margins of this limit
($\beta_S \approx 5.5$) so that if relativistic
(or dissipative) effects drive the WD too far inward from the initial Keplerian
trajectory, much of the tidal debris can be swallowed directly and leave little
unbound material. However the parameter space
associated with $10^4 M_\odot$ BHs that is also capable of giving rise to robust nucleosynthesis is extremely limited.
So we consider this case only to gauge the effect of BH mass and tidal force gradient on 
the efficiency of nucleosynthesis, even if most of the debris is bound to or swallowed by the BH. For this reason we restrict
calculations of this binary system to just two levels of refinement. 
A few of the more interesting case A scenarios are run at both two and three
refinement levels to assess convergence.

These relatively moderate values of $\beta$ conveniently place the perihelion radii of all spin-orbit 
orientations that we consider outside of the inner most stable circular orbit (ISCO) and the marginally bound orbit (MBO)
\begin{equation}
r_{\mathrm{MB}} = 2M \pm a + 2 \sqrt{M (M \pm a) }   ~,
\end{equation}
where the + (-) sign applies to retrograde (prograde) orbits.
The stars have radii smaller than the difference between perihelion and ISCO radii, suggesting that
very little WD material will get swallowed by the BH prior to passing through periapsis,
assuming the interaction strength is also sufficiently smaller than $\beta_S$.
For reference, the (ISCO, MBO) radii evaluate to (2.3, 1.7) $R_G$ for prograde and
(8.7, 5.7) $R_G$ for retrograde orbits around a
BH with spin parameter $a=0.9 M_\mathrm{BH}$. Of course these degenerate to (6, 4) $R_G$ with zero spin.

\begin{figure}
\includegraphics[width=0.5\textwidth]{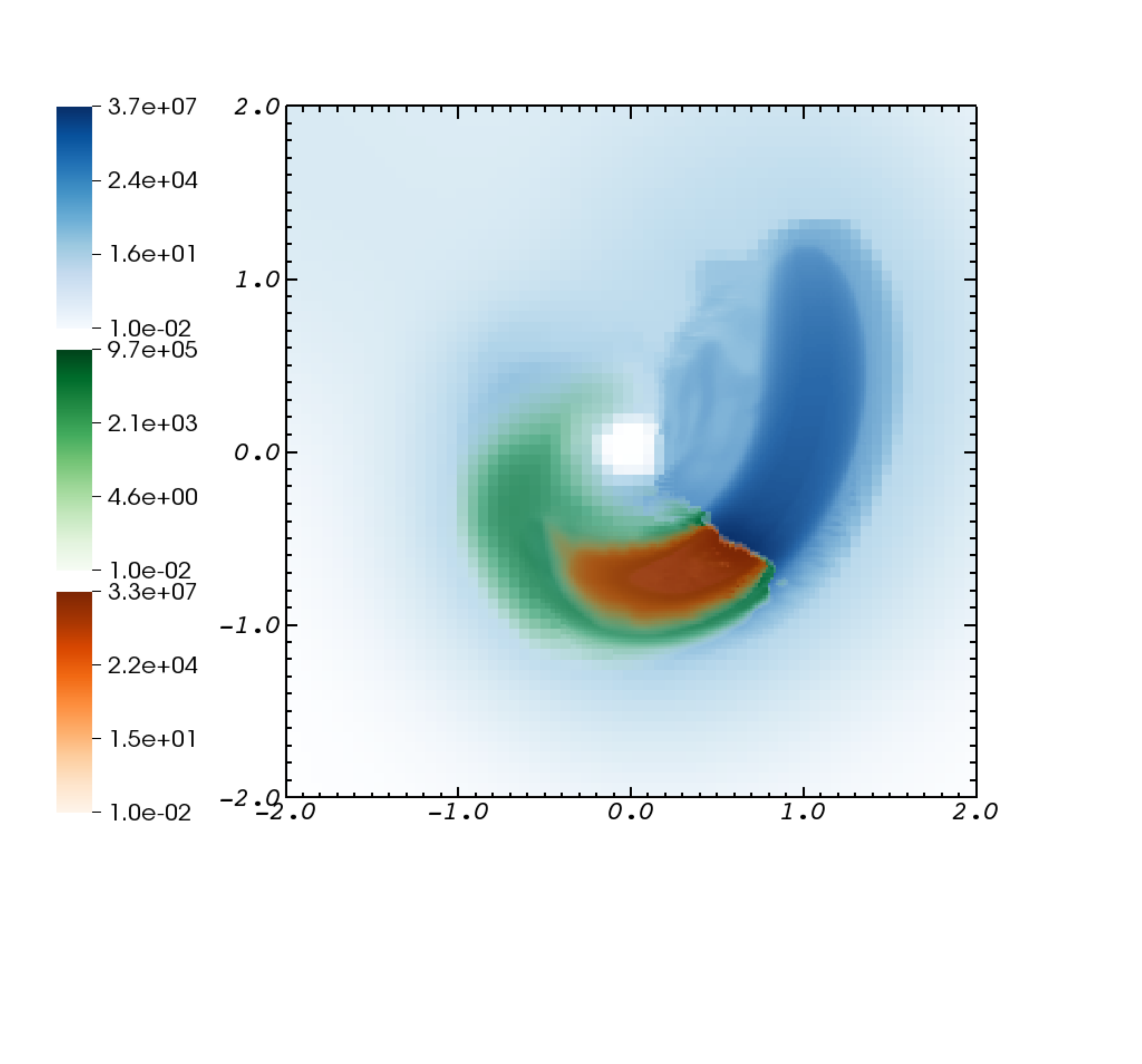}
\includegraphics[width=0.5\textwidth]{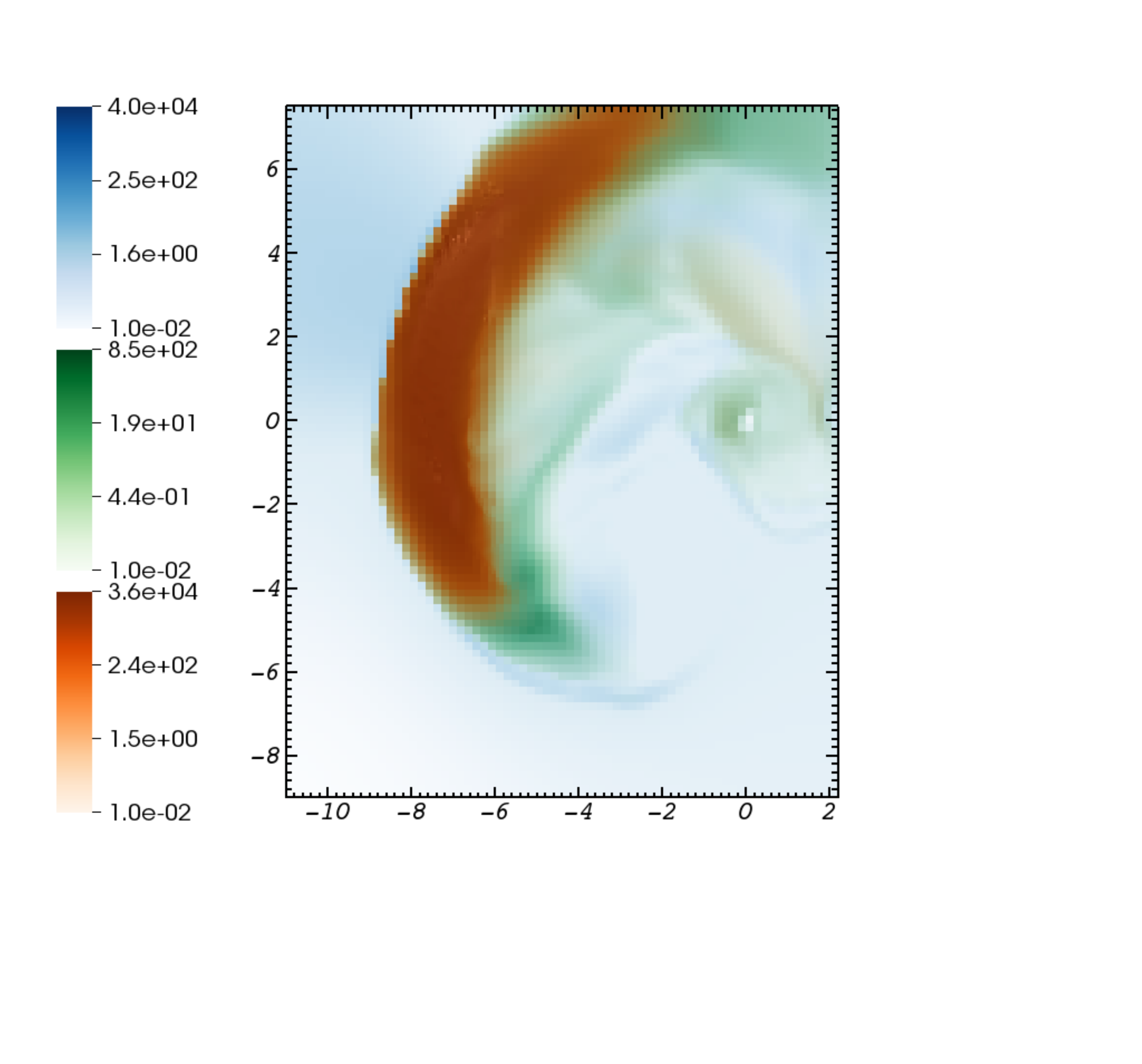} \\
\includegraphics[width=0.5\textwidth]{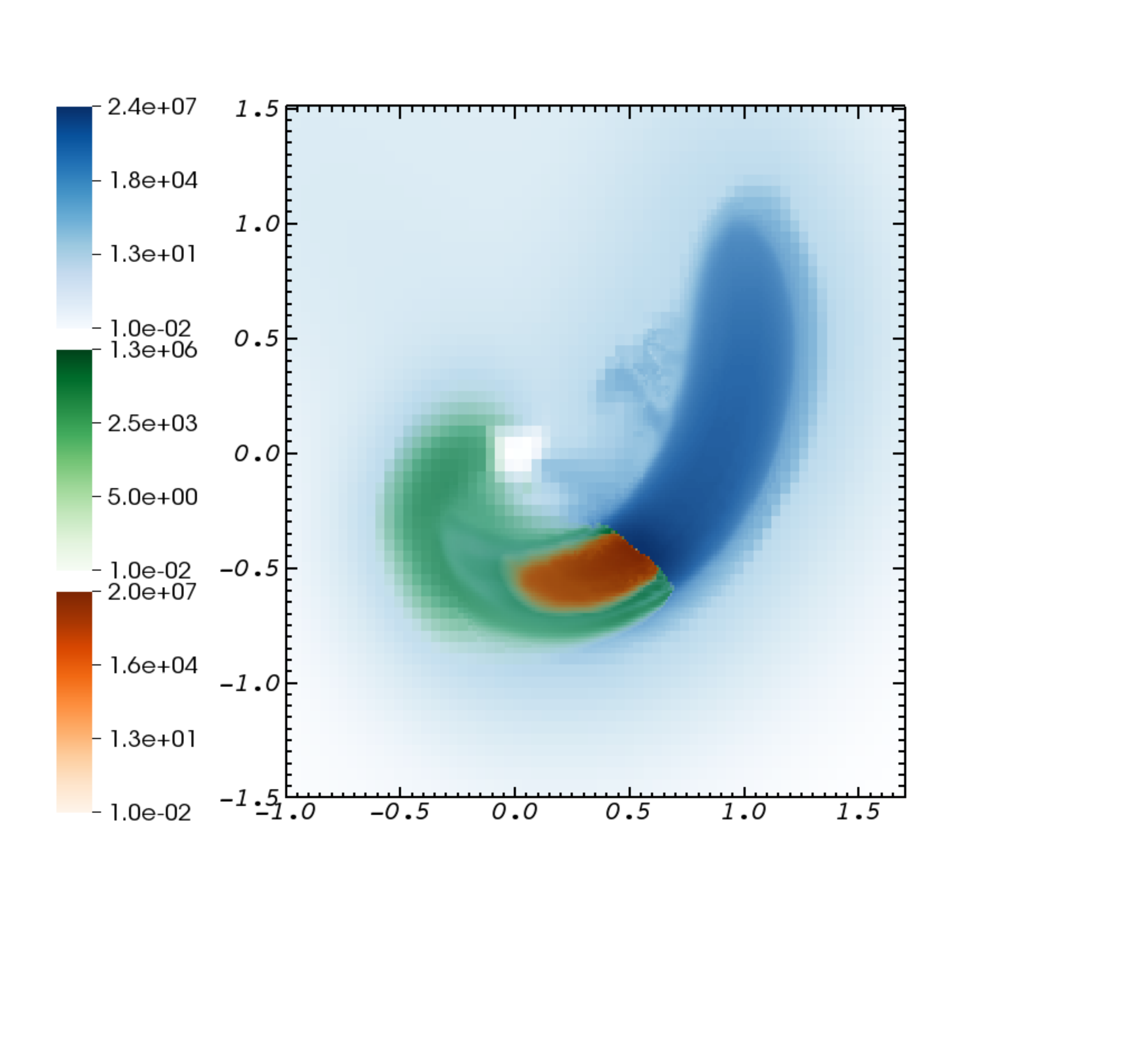}
\includegraphics[width=0.5\textwidth]{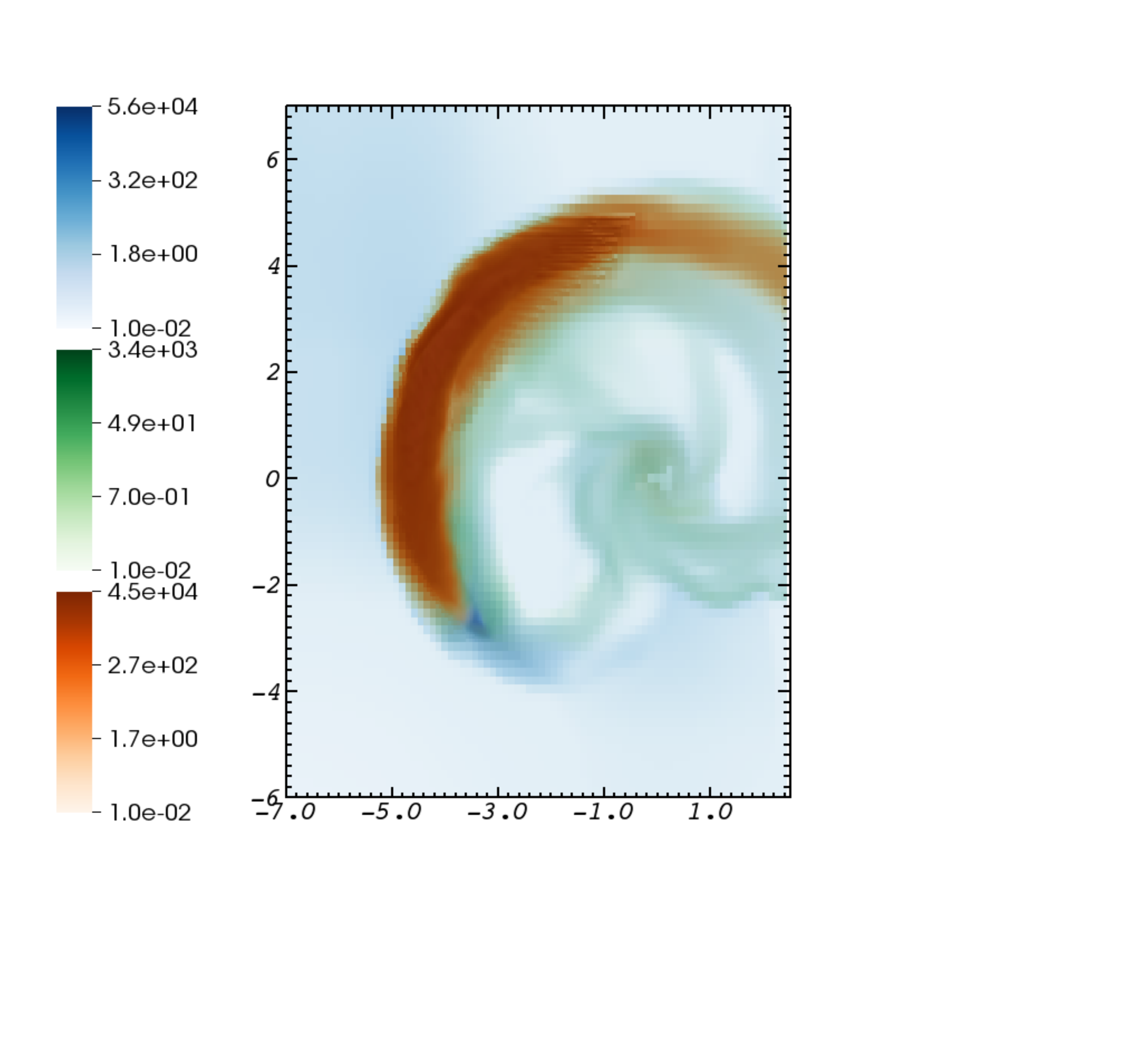}
\caption{
Logarithm of the $^{56}$Ni (orange color map), $^{40}$Ca (green color map), and total mass densities (blue color map) 
sliced through the orbital plane for case A9p-c at times 0.04 (top left) and 2 secs (top right), and
for case A9r at 0.02 (bottom left) and 2 secs (bottom right).
Legends are densities in g cm$^{-3}$. Spatial scales are in units of $R_P$.
}
\label{fig:images_den}
\end{figure}

\begin{figure}
\includegraphics[width=0.5\textwidth]{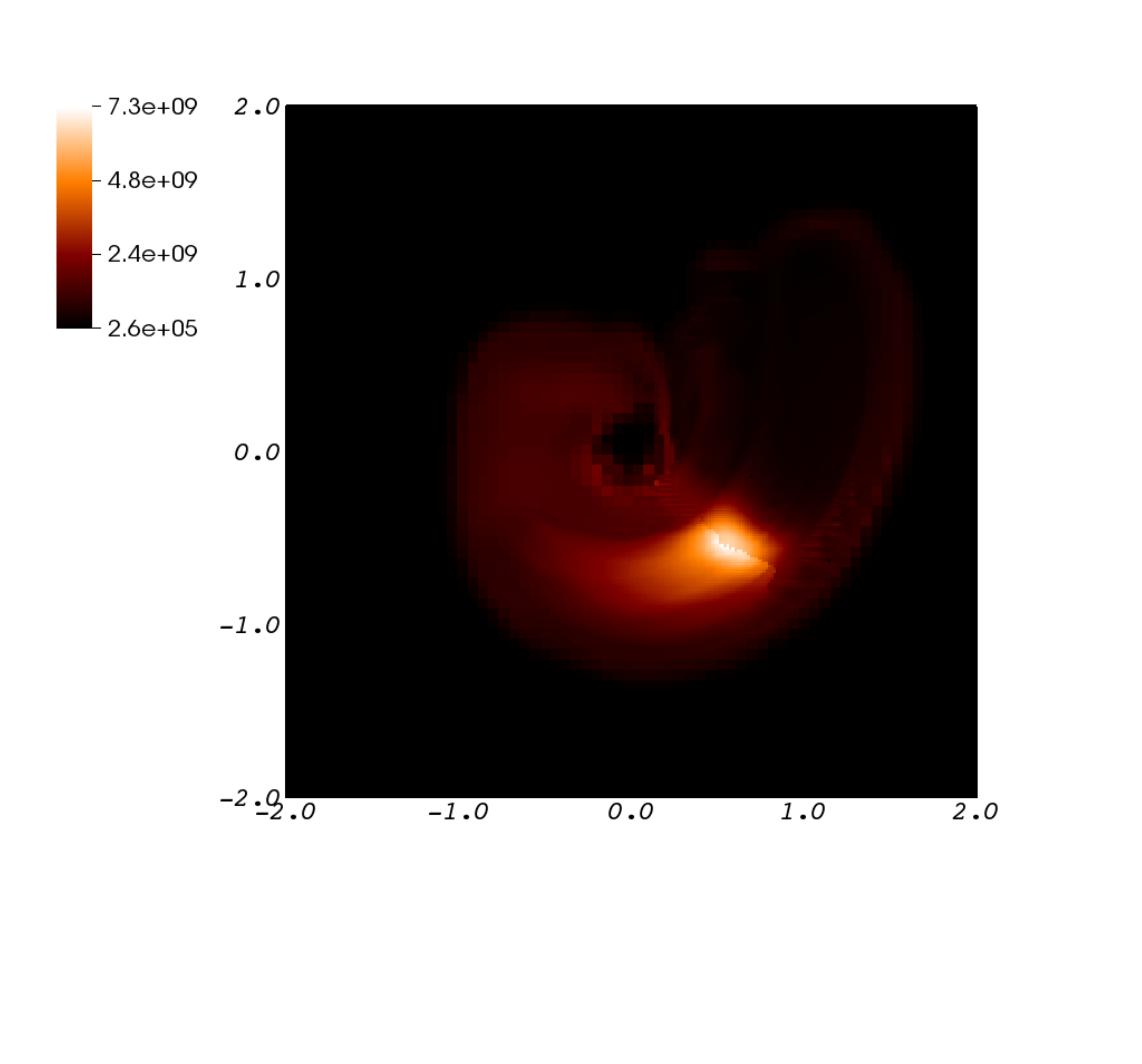}
\includegraphics[width=0.5\textwidth]{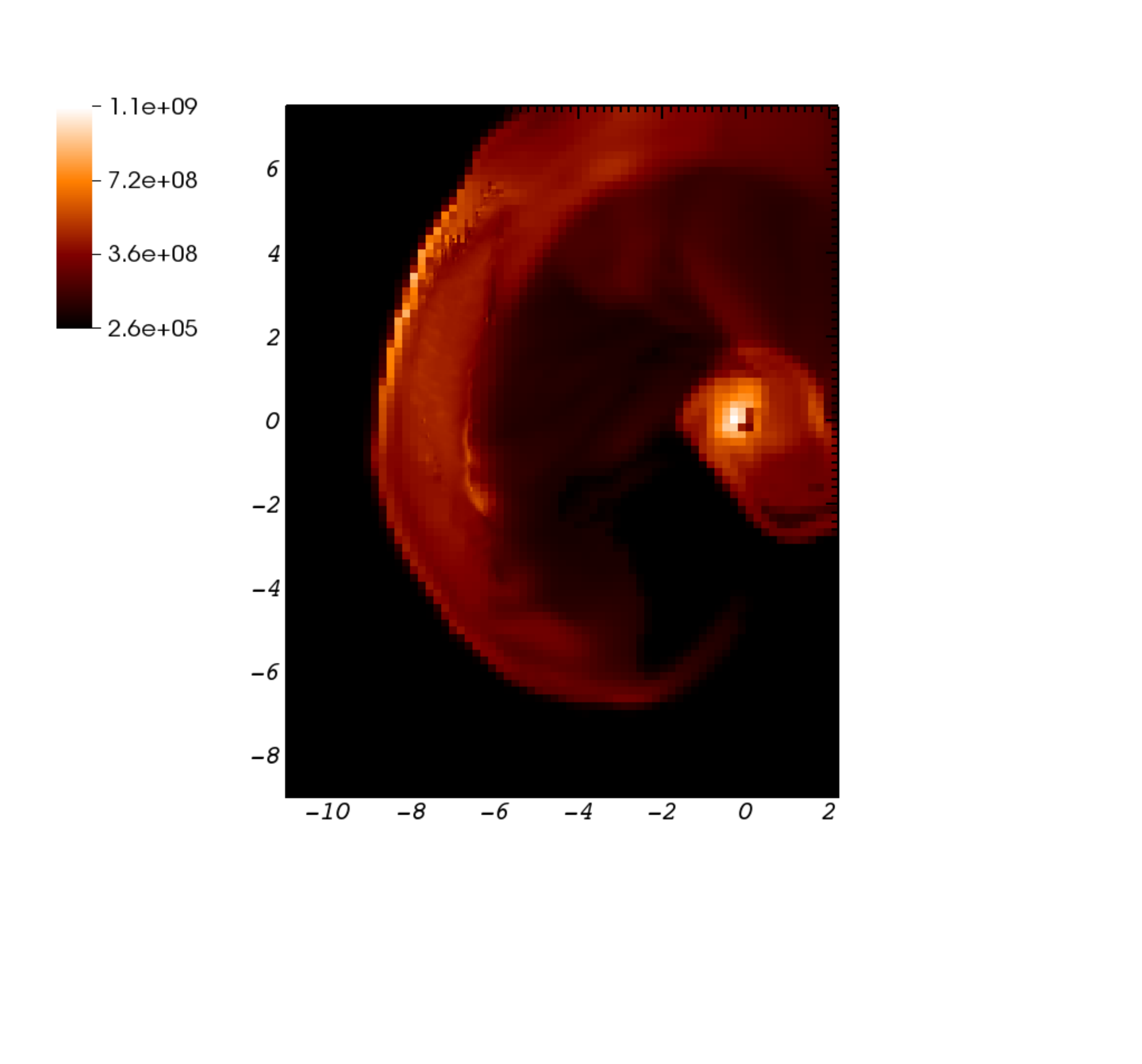} \\
\includegraphics[width=0.5\textwidth]{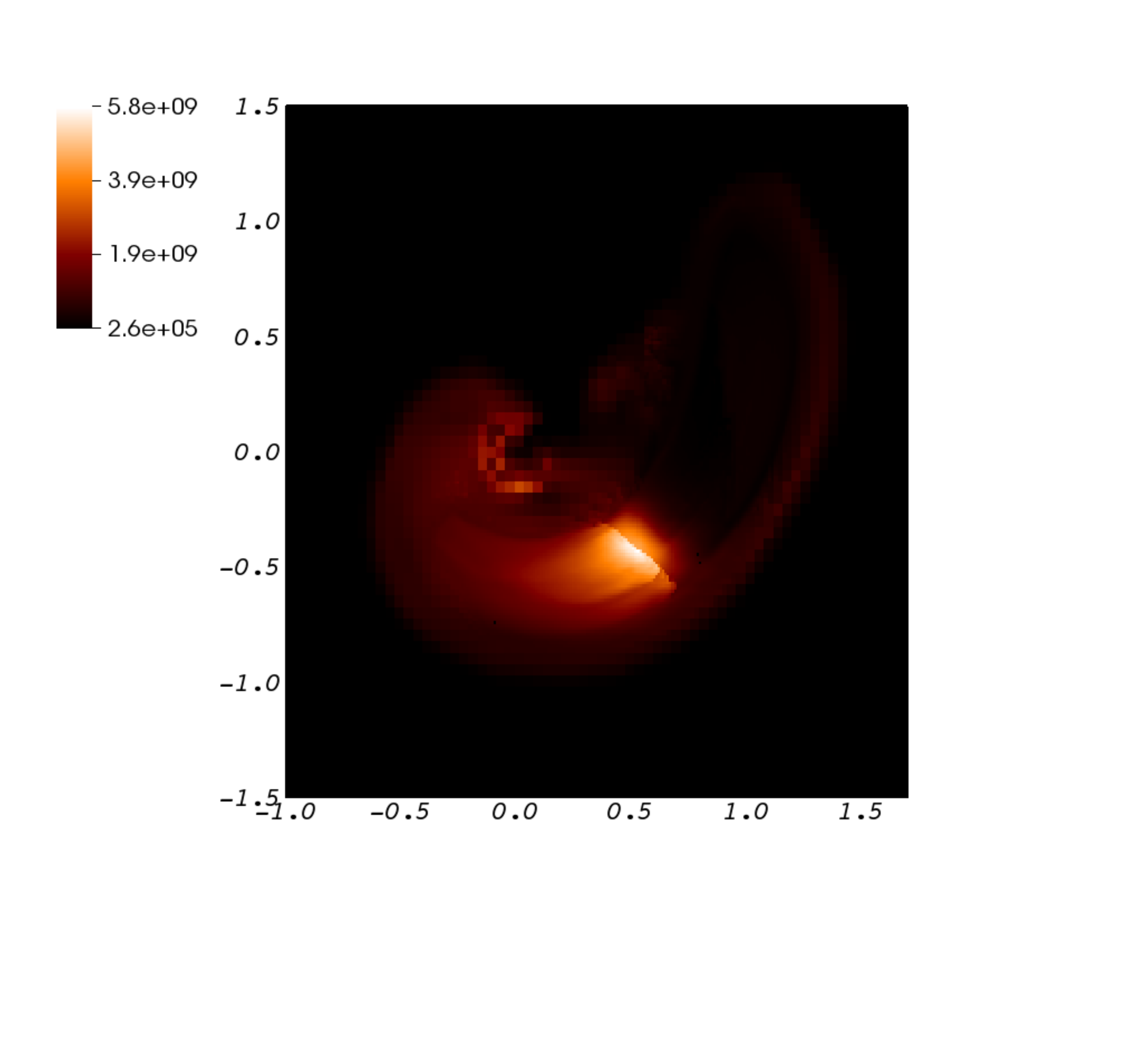}
\includegraphics[width=0.5\textwidth]{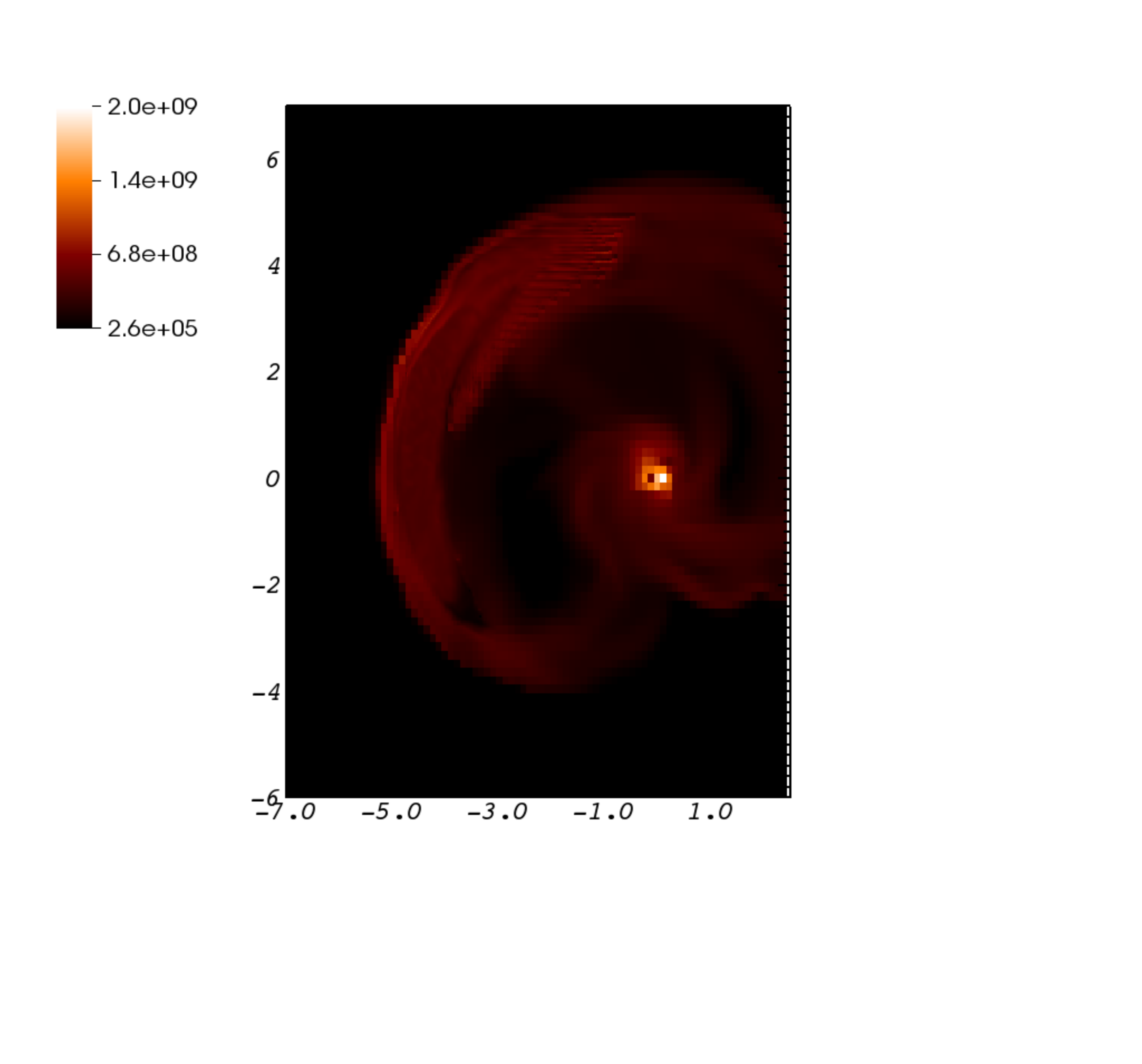}
\caption{
Gas temperature in units of Kelvin corresponding to the image sequences of Figure \ref{fig:images_den}.
}
\label{fig:images_tem}
\end{figure}

\begin{figure}
\includegraphics[width=0.8\textwidth]{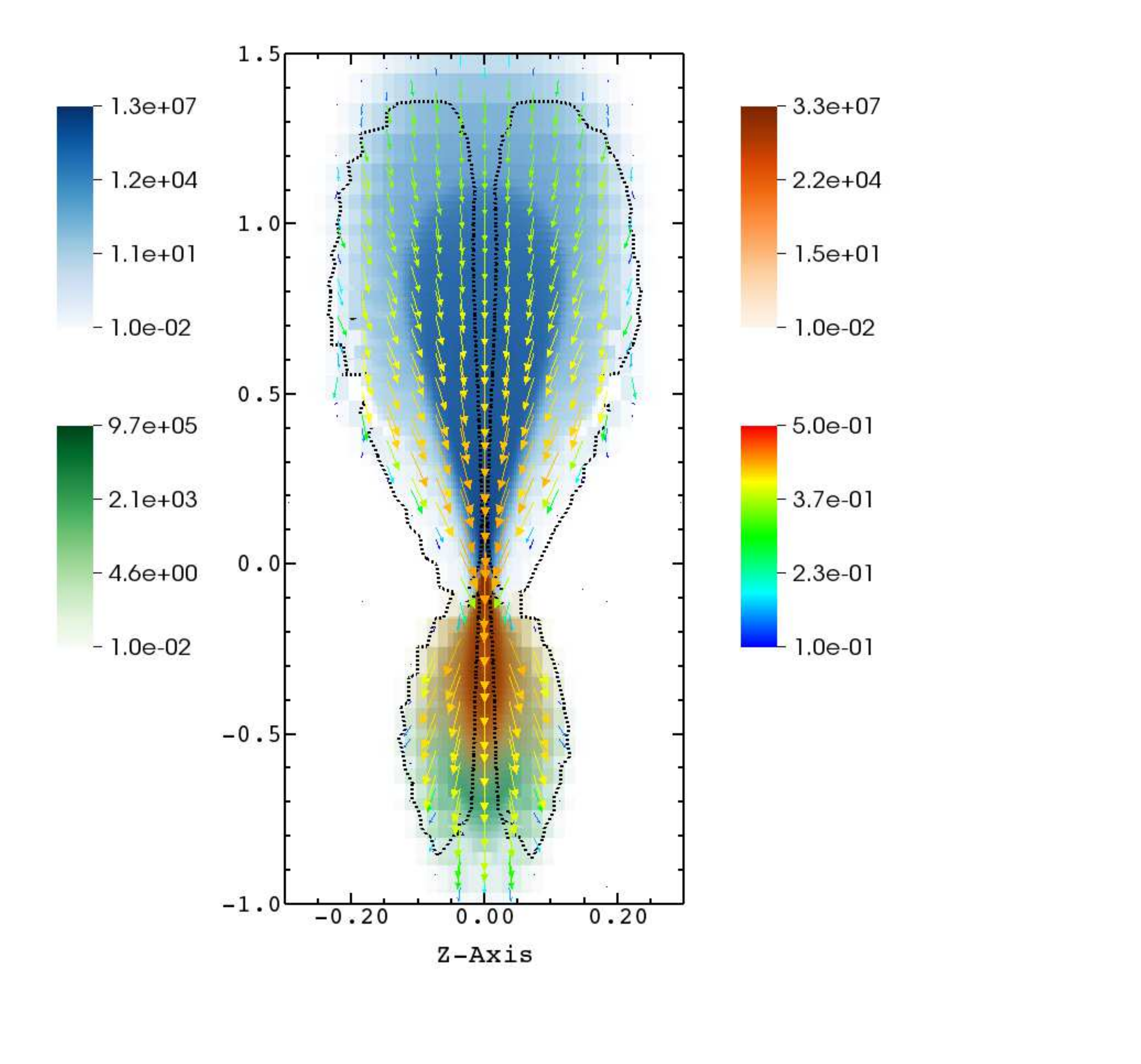}
\caption{
Logarithm of the $^{56}$Ni (orange color map), $^{40}$Ca (green), and $^{12}$C mass densities (blue) 
sliced through the center of the hotspot developing in Figure \ref{fig:images_den}
in a plane orthogonal to the line directed towards the black hole.
The image corresponds to case A9p-c at 0.04 secs, the same time as the left image of Figure \ref{fig:images_den}.
Vectors (rainbow map) track the fluid motion compressing towards the $z=0$ orbital plane (top lobe), then 
decompressing supersonically (bottom lobe) after bouncing off the mid-plane. The dotted black line
is the contour surface where the Mach number of the z-velocity component is unity.
Densities are in g cm$^{-3}$, velocities are scaled to the speed of light,
and axes labels are in units of $R_P$.
}
\label{fig:images_z}
\end{figure}

Figure \ref{fig:images_den} shows color images of the logarithm of the $^{56}$Ni, $^{40}$Ca, and
total (sum of all species) mass densities in the orbital plane for cases A9p-c and A9r at early and late times
when the WD first approaches periapsis and as it ends its first clockwise pass around the black hole.
Figure \ref{fig:images_tem} shows the corresponding gas temperatures in units of Kelvin.
These images exhibit typical behaviors found in all of our encounter scenarios: 
Calcium group elements begin to form the moment
stellar matter compresses to maximum density in the orbital plane and heats
to ignition temperatures, primarily along the radially aligned feature evident in the early
(left) images of both the density and temperature figure sets. 
Shocks develop above and below the orbital plane, as shown in Figure \ref{fig:images_z},
contributing to the ignition of nuclear fuel and thermal runaway conditions
capable of burning through to iron. Burn products, after having formed in the
central high density regions, follow in the wake of these shocks
which move off the orbital plane at velocities between 0.03 to $0.09 c$.
Heavier elements subsequently form within the confines of the lighter elements,
giving rise to a layered structure resembling SN Ia models where the densest
parts of the debris are composed mostly of iron products surrounded
by intermediate mass elements, which in turn are embedded within an outer layer of unspent fuel.
Once nucleosynthesis gets started, the composition of burn products is generally established
fairly quickly, within a half second or so before stellar material decompresses.

Although highly resolved for three dimensions, our calculations do not cover fully the 
spatial range needed to determine precisely where the shocks emerge, or whether they are
responsible for triggering or simply enhancing ignition. Instead we refer to the work of
\citet{Tanikawa17} who performed much more highly resolved 1D simulations down to scales
of $10^4$ cm, adopting dynamical flow conditions from 3D SPH models. Their calculations
demonstrate a network of shocks develop within the scale height of stellar matter at different
locations above and below the orbital plane. The forward-most shocks (which in the 3D
calculations are delineated by the outer unit
Mach number contour in the lower lobe of Figure \ref{fig:images_z}) do not accompany sustained
detonation waves; they decay due to propagating through low density fuel.
Sustained detonations occur within a few hundred kilometers from the orbital plane
(near the inner Mach number contour of Figure \ref{fig:images_z})
and are powered by a combination of tidal compression and shocks so long
as they break out at densities $\gtrsim 10^6$ g cm$^{-3}$.
Their work additionally established a minimum
resolution requirement for 3D calculations of $\lesssim 10^6$ cm in order to achieve
reasonably converged solutions for triggering detonation and generating nuclear yields, 
a scale that we have intentionally matched in this work.

Our calculations suggest detonations may be triggered spontaneously, not by direct
initiation. Although we cannot entirely rule out the latter possibility due to
limited resolution, we find no evidence in the simulations that hydrodynamic shocks
trigger ignition. Rather they appear to enhance developing hotspots and to facilitate
thermonuclear burn after the fuel first ignites from induction gradients within the
shock confined regions. In this scenario it is important to resolve
the critical length scale needed to generate a detonation wave, which rapidly
decreases with increasing fuel density. Tracing the origin of the initiation, we find pockets of 
\replaced{high nickel production}{burn products} at the earliest stages of nucleosynthesis correlate with 
\replaced{the central densest portions of the WD.}{the highest density regions at the center of the WD.}
These regions host roughly two-to-one
oxygen/carbon ratios \added{at mean densities $\sim 10^7$ g cm$^{-3}$ and temperatures between
$2 - 2.5\times10^9$ K, slightly less dense and cooler than their peak values at maximum compression.}
For such conditions Table 11 from \citet{Seitenzahl09}
predicts a minimum critical length scale of about 50 km, a number that they estimate
is likely a factor of two or three low based on their resolution studies and well
above \replaced{the}{our} 10 km cell size.

\subsection{Inline Nucleosynthesis}
\label{subsec:nucleoresults}

Table \ref{tab:results} summarizes our results at both high and low grid resolutions.
Displayed are the actual (relativistically corrected) tidal strength, maximum densities, maximum temperatures, global nuclear energy
production, and final (plateau) calcium and iron-group masses. 
We define iron group as those elements with atomic numbers in the range 52-60,
calcium group between 40-51, and silicon between 28-39.
For the 19-isotope inline network these group elements consist simply of
(Fe, Ni), (Ca, Ti, Cr), and (Si, S, Ar), respectively.

\begin{deluxetable}{lcccccc}
\tablecaption{Inline Result Summary \label{tab:results}}
\tablewidth{0pt}
\tablehead{
\colhead{Run}                     & 
\colhead{$\beta^*$}               & 
\colhead{$\rho_\mathrm{max}$}     & 
\colhead{$T_\mathrm{max}$}        &
\colhead{$e_\mathrm{nuc}$}        & 
\colhead{$M_\mathrm{Fe, max}$}    & 
\colhead{$M_\mathrm{Ca, max}$}    \\
                                  & 
                                  & 
(g cm$^{-3}$)                     & 
(K)                               &
(erg)                             & 
($M_\odot$)                       & 
($M_\odot$)
}
\startdata
A9p-c      & 8.5 & $2.0\times10^7$ & $5.6\times10^9$ & $7.5\times10^{50}$ & $2.7\times10^{-1}$ & $2.1\times10^{-2}$ \\
A9p-b      & 6.3 & $1.8\times10^7$ & $4.8\times10^9$ & $6.2\times10^{50}$ & $1.2\times10^{-1}$ & $1.4\times10^{-2}$ \\
A0-b       & 6.4 & $1.7\times10^7$ & $4.6\times10^9$ & $5.8\times10^{50}$ & $8.8\times10^{-2}$ & $1.6\times10^{-2}$ \\
A9r        & 6.3 & $1.8\times10^7$ & $4.5\times10^9$ & $5.8\times10^{50}$ & $9.1\times10^{-2}$ & $1.5\times10^{-2}$ \\
A0-a       & 5.5 & $1.6\times10^7$ & $4.2\times10^9$ & $4.9\times10^{50}$ & $4.4\times10^{-2}$ & $1.0\times10^{-2}$ \\
A9p-a      & 5.1 & $1.6\times10^7$ & $4.0\times10^9$ & $4.4\times10^{50}$ & $3.3\times10^{-2}$ & $8.0\times10^{-3}$ \\
A9m        & 5.5 & $1.2\times10^7$ & $3.9\times10^9$ & $4.6\times10^{50}$ & $3.5\times10^{-2}$ & $7.5\times10^{-3}$ \\
\\
{\bf A9p-c}& 8.4 & $4.6\times10^7$ & $6.8\times10^9$ & $8.3\times10^{50}$ & $4.0\times10^{-1}$ & $1.8\times10^{-2}$ \\
{\bf A9r}  & 6.3 & $3.5\times10^7$ & $5.3\times10^9$ & $7.3\times10^{50}$ & $2.1\times10^{-1}$ & $2.3\times10^{-2}$ \\
{\bf A9p-a}& 5.1 & $2.8\times10^7$ & $4.6\times10^9$ & $6.0\times10^{50}$ & $9.0\times10^{-2}$ & $2.1\times10^{-2}$ \\
\\
B9p-b      & 7.0 & $5.0\times10^5$ & $1.5\times10^9$ & $6.1\times10^{49}$ & $1.3\times10^{-4}$ & $1.1\times10^{-2}$ \\
B0-b       & 7.7 & $5.3\times10^5$ & $1.4\times10^9$ & $5.5\times10^{49}$ & $9.3\times10^{-5}$ & $9.1\times10^{-3}$ \\
B9r        & 8.3 & $5.9\times10^5$ & $1.0\times10^9$ & $3.2\times10^{49}$ & $1.4\times10^{-5}$ & $4.3\times10^{-3}$ \\
B0-a       & 5.4 & $5.0\times10^5$ & $7.1\times10^8$ & $6.3\times10^{48}$ & $ <10^{-15}$ & $5.0\times10^{-12}$ \\
B9m        & 5.3 & $3.3\times10^5$ & $5.4\times10^8$ & $2.5\times10^{48}$ & $ <10^{-15}$ & $ <10^{-15}$ \\
B9p-a      & 4.7 & $4.5\times10^5$ & $4.1\times10^8$ & $3.4\times10^{47}$ & $ <10^{-15}$ & $ <10^{-15}$ \\
\enddata
\end{deluxetable}

There is generally good agreement in the density/temperature phase tracks between low and high resolution cases:
temperatures match to roughly 20\%, and the densities to about a factor of two,
suggesting that we have come close to converging on the mid-plane compression, but not quite.
These differences manifest as 30\% uncertainty in the total nuclear energy released, and
about a factor of two in burn product synthesis.

It is not surprising to find some sensitivity of our results to grid resolution since the critical
length scales for carbon and helium burn are small and difficult to
achieve in three dimensional calculations \citep{Seitenzahl09,Holcomb13}.
However it is encouraging that our results are not strongly dependent on resolution, as would be the
case if they are entirely due to poor resolution. As noted earlier,
we attribute this to matching the grid scale conditions ($\lesssim 10^6$ cm) suggested by the 1D
studies carried out by \citet{Tanikawa17} for establishing converged solutions
at the interaction strengths considered in this work.

Comparing the two nonrotating cases A0-a and B0-a to the nearest equivalent cases in Paper 1
we can additionally assess effects from all the physics upgrades incorporated into these new studies, namely the
{\sc MESA} (as opposed to TOV) generated stellar models, the Helmholtz EOS, and dynamic self-gravity. 
Because of the different stellar models, there is no exact
encounter scenario to which we can directly compare results. However we point out that
case A0-a is close in interaction strength to B3M6R12 from Paper 1, but closer
to B3M6R9 in perihelion approach to the BH. Comparing our current results to both of these cases
we find generally good agreement in the peak density, peak temperature, total nuclear energy 
and calcium production, roughly an average factor of two in all these diagnostics.
Iron production differs the most, but our current results lie between these two cases
which happen to straddle the parameter space in a region of strong sensitivity.

The nonrotating case B0-a has an interaction strength
and perihelion radius slightly greater than but comparable to B4M2R6 from Paper 1. These two cases
similarly agree quite well in peak density and temperature (about a factor of 2),
energy released (10\%), and small or negligible levels of both calcium and iron group elements.
Our previous results presented in Paper 1 are thus validated to 
roughly a factor of 2 collectively, despite using less sophisticated physical models.

Quantifying nuclear activity by both the nuclear energy released and mass conversion efficiency to iron products,
we find that rotation effects from $10^3 M_\odot$ BH encounters is quite small. Adjusting the perihelion radius
to produce comparable $\beta^* \approx 5.5$ interaction strengths, nuclear activity varies by only 25\%
comparing cases A0-a (no rotation), A9m ($a=0.9$, polar orbit), and A9p-a ($a=0.9$, prograde orbit).
A similarly small variation is observed comparing $\beta^* \approx 6$ encounters
with a $10^3 M_\odot$ BH (cases A0-b, A9r, and A9p-b), and $\beta^* \approx 8$ encounters
with a $10^4 M_\odot$ BH (cases B0-b, B9r, and B9p-b), regardless of whether the BH is
rotating or not, or whether the spin-orbital vectors are aligned or not.
These comparisons are evaluated at the relativistic
(not Newtonian) penetration factors $\beta^*$ and thus account
for the actual tidal strength experienced by the stellar mass centroids in each of the models,
which depends rather sensitively on BH spin alignment: 
retrograde (prograde) orbits approach closer to (further from) the BH than the Keplerian approximation would predict due
to the loss (gain) of angular momentum from frame dragging effects.
Whether this small difference is physical or numerical is difficult to say: 
as we have noted above, 25\% variance in nuclear energy is roughly the level of numerical uncertainty at our current grid resolution.

\subsection{Post-processed Nucleosynthesis}
\label{subsec:postresults}

At the start of each calculation, stellar interiors are sampled with approximately fifteen-thousand particles
that are evolved (passively advected) in time according to their Lagrangian velocity using a second order
predictor-corrector scheme.
The initial angular distribution of particles inside the white dwarf is random but isotropic.
Radial positions are randomly sampled from a mass distribution function derived
by inverting the {\sc MESA} generated mass-radius profiles, which produces a nearly uniform mass for all
particles and concentrates particles towards the core center in a way that respects the stellar density profile.
All particles are endowed with physical attributes (e.g., temperature, density, and energy density)
that are interpolated using an inverse covariant distance weighting scheme contributed to by
nearest (shared face) neighbor cells and projected to particle positions.
Particle positions and thermodynamic states are computed each cycle as part of the hydrodynamic update sequence, but
are stored to disk at less regular intervals in preparation for post-processing with a more elaborate
495-isotope nuclear reaction network adapted from the Torch code \citep{Timmes99}.
Particles falling into the BH horizon are discarded and thus ignored in the final analysis.
This treatment improves on the post-processing method of Paper 1 by sampling the entire star (not just the
inner dense core), and by tracking the stellar interior along Lagrangian fluid elements 
(not just within the densest Eulerian cells).

The specific energy spread associated with tidal forces is approximated to within
a numerical constant of order unity as \citep{Stone13,Kawana18}
\begin{equation}
\Delta\epsilon_{\mathrm{tidal}} \approx \frac{G M_{\mathrm{BH}} R_{\mathrm{WD}}}{R_T^2} 
                       \approx 1.2\times10^{-3} c^2
                        \left(\frac{R_\mathrm{WD}}{10^9 \mathrm{cm}}\right)^{-1}
                        \left(\frac{M_\mathrm{BH}}{10^3 M_\odot}\right)^{1/3}
                        \left(\frac{M_\mathrm{WD}}{0.6 M_\odot}\right)^{2/3} ~.
\label{eqn:spread}
\end{equation}
Considering lower mass stars also have larger radii, two of the three binary system parameters
in equation (\ref{eqn:spread}) conspire to produce smaller energy spreads for the B-series calculations.
Additionally, the relatively inefficient nuclear energy released
(in Table \ref{tab:results}) from this class B encounter scenario suggests that the effect of nuclear reactions
on debris motion is small since $\Delta\epsilon_{\mathrm{tidal}} > \Delta\epsilon_{\mathrm{nuc}}$, where
$\Delta\epsilon_{\mathrm{nuc}} \approx 10^{49}\mathrm{erg}/M_\mathrm{WD} \approx 10^{-4} c^2$.
Hence in this section we consider only the A-series calculations where nuclear energy release
is expected to be more important.

\begin{figure}
\includegraphics[width=0.6\textwidth]{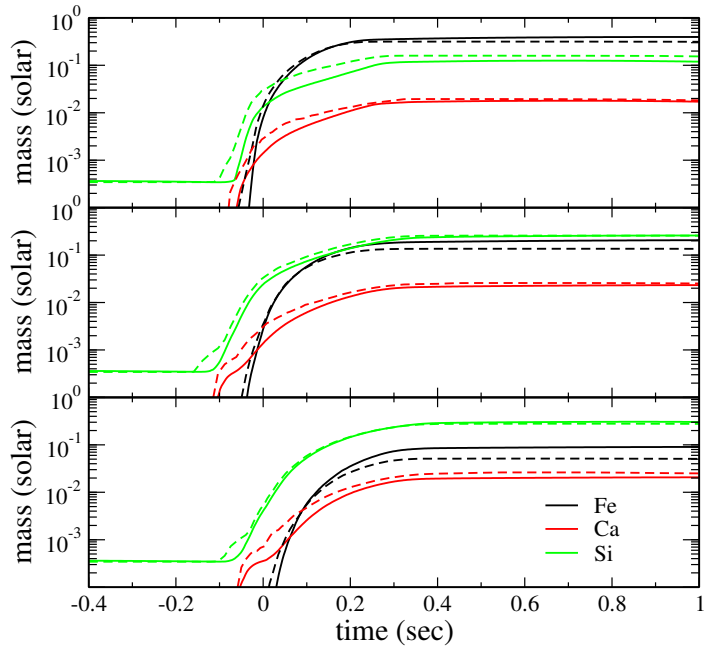}
\caption{
Production and evolution of Fe, Ca, and Si group elements from the high 
resolution cases: A9p-c (top), A9r (middle), and A9p-a (bottom). Solid (dotted) lines
represent results from the 19 (495) isotope network model.
}
\label{fig:pp_mass_all}
\end{figure}

Table \ref{tab:pp_results} summarizes the total nuclear energy released, along with the peak calcium
($M_\mathrm{Ca,max}$) and iron ($M_\mathrm{Fe,max}$)
masses produced in each calculation, which should be compared (and be comparable) to the corresponding
results in Table \ref{tab:results} calculated with the inline 19-isotope model.
Looking especially at the high resolution calculations, we find the 19 and 495 isotope models 
agree remarkably well; roughly 20\% on the plateau (final) total energy release and calcium production, and 
within 75\% on the amount of iron. The production histories of
Fe, Ca, and Si group elements are plotted in Figure \ref{fig:pp_mass_all} for the high resolution
cases A9p-c, A9r, and A9p-a. Solid (dotted) lines represent results from
the inline 19 (post-processed 495) isotope network models. 
Apart from the relative amounts of Fe, Ca, and Si produced, which transitions from Fe dominated to
Si dominated with increasing perihelion radius,
these results are typical in that the more accurate 495 isotope model predicts
slightly greater intermediate element production at the expense of iron. Agreement
between the simple and complex models is otherwise excellent for all group elements and throughout time.

\begin{deluxetable}{lcccccccc}
\tablecaption{Post-processed Result Summary \label{tab:pp_results}}
\tablewidth{0pt}
\tablehead{
\colhead{Run}                     & 
\colhead{$\beta^*$}               & 
\colhead{$e_\mathrm{nuc}$}        & 
\colhead{$M_\mathrm{tot,unbnd}$}  & 
\colhead{$M_\mathrm{Fe, max}$}    & 
\colhead{$M_\mathrm{Fe, unbnd}$}  & 
\colhead{$M_\mathrm{Ca, max}$}    & 
\colhead{$M_\mathrm{Ca, unbnd}$}  \\
                                  & 
                                  & 
(erg)                             & 
($M_\odot$)                       & 
($M_\odot$)                       & 
($M_\odot$)                       & 
($M_\odot$)                       & 
($M_\odot$)
}
\startdata
A9p-c    & 8.5 & $7.5\times10^{50}$ & $3.0\times10^{-1}$ & $2.5\times10^{-1}$ & $1.3\times10^{-1}$ & $2.4\times10^{-2}$ & $1.1\times10^{-2}$ \\
A9p-b    & 6.3 & $6.4\times10^{50}$ & $3.0\times10^{-1}$ & $1.0\times10^{-1}$ & $6.1\times10^{-2}$ & $3.0\times10^{-2}$ & $1.3\times10^{-2}$ \\
A0-b     & 6.4 & $6.1\times10^{50}$ & $2.8\times10^{-1}$ & $7.2\times10^{-2}$ & $3.7\times10^{-2}$ & $2.8\times10^{-2}$ & $1.0\times10^{-2}$ \\
A9r      & 6.3 & $5.8\times10^{50}$ & $3.0\times10^{-1}$ & $6.2\times10^{-2}$ & $3.6\times10^{-2}$ & $2.6\times10^{-2}$ & $1.1\times10^{-2}$ \\
A0-a     & 5.5 & $5.1\times10^{50}$ & $2.7\times10^{-1}$ & $2.5\times10^{-2}$ & $1.3\times10^{-2}$ & $2.3\times10^{-2}$ & $8.2\times10^{-3}$ \\
A9p-a    & 5.1 & $4.7\times10^{50}$ & $2.9\times10^{-1}$ & $1.8\times10^{-2}$ & $1.2\times10^{-2}$ & $2.0\times10^{-2}$ & $8.1\times10^{-3}$ \\
A9m      & 5.5 & $5.1\times10^{50}$ & $3.2\times10^{-1}$ & $3.0\times10^{-2}$ & $1.7\times10^{-2}$ & $2.2\times10^{-2}$ & $1.0\times10^{-2}$ \\
\\
{\bf A9p-c}& 8.4 & $7.1\times10^{50}$ & $3.0\times10^{-1}$ & $3.2\times10^{-1}$ & $2.0\times10^{-1}$ & $1.9\times10^{-2}$ & $7.4\times10^{-3}$ \\
{\bf A9r}  & 6.3 & $6.0\times10^{50}$ & $2.7\times10^{-1}$ & $1.4\times10^{-1}$ & $8.4\times10^{-2}$ & $2.6\times10^{-2}$ & $1.1\times10^{-2}$ \\
{\bf A9p-a}& 5.1 & $5.0\times10^{50}$ & $2.6\times10^{-1}$ & $5.1\times10^{-2}$ & $3.4\times10^{-2}$ & $2.6\times10^{-2}$ & $1.1\times10^{-2}$ \\
\enddata
\end{deluxetable}

Also shown in Table \ref{tab:pp_results} is the total unbound mass ($M_\mathrm{tot,unbnd}$) and the unbound mass of burn products
($M_\mathrm{Fe,unbnd}$, $M_\mathrm{Ca,unbnd}$) estimated by summing up the mass of all particles found with positive
specific orbital energy ($\epsilon=v^2/2 - GM/r$) at the final simulation time of 2 seconds past periapsis.
At interaction strength $\beta^* \approx 5.5$ (runs A0-a, A9m, and A9p-a), the average mass conversion efficiency of
total (bound plus unbound) iron and calcium is roughly the same, 3-5\%, with a very weak (if any) dependence on BH spin.
About 40-50\% of all iron and 30-40\% of all the calcium produced is unbound in these cases, regardless of BH spin.

Prograde orbital alignment allows for the WD companion to approach closer to the BH without falling
inside the ISCO (or MBO), offering the intriguing possibility of producing more unbound intermediate and heavy
elements. As expected, we do observe significantly greater iron production
at smaller perihelion radii (compare cases A9p-a, A9p-b, and A9p-c, where the latter case reduces
the perihelion radius to the minimum value before the outer layers of WD material hit the ISCO). However the
capture cross-section increases as well and we find a somewhat smaller, but still an order of magnitude increase, in the
amount of unbound iron group elements. The total (unbound) mass conversion efficiency for these scenarios peaks at
about 50\% (70\%) in the high resolution case A9p-c.
Roughly 50-60\% of all iron and 35-40\% of all calcium group elements are unbound in both
the more distant (A9p-a) and closer (A9p-c) cases, though the latter produces an
order of magnitude more unbound and total iron mass.
Furthermore, whereas the fraction of unbound iron increases with decreasing perihelion
radius, the fraction of unbound calcium remains nearly constant, suggesting 
(as we demonstrated in Paper 1) that the ratio
of unbound intermediate to heavy group elements might prove to be a defining signature of binary
system parameters.
Comparing cases A9p-a and A9p-c where the perihelion radius is reduced by about 1/3, the ratio 
of unbound calcium to unbound iron mass is reduced by an order of magnitude.

By way of validation, we point out that our results concerning the amount of unbound debris are
generally consistent with those published by \citet{Kawana18} for the case of nonrotating BHs.
We additionally note that the rotating BH case A9p-c evolves to a comparable penetration factor as run B6Su in \citet{Haas12},
and results in a nearly identical total unbound mass fraction, 50\% compared to 60\%.

\subsection{Spatial Distribution}
\label{subsec:distribution}

All calculations are terminated shortly after nucleosynthesis settles to small fractional
energy changes, no longer than 2 seconds after reaching periapsis. Hence we do not
(in this work) follow fall-back effects. Instead we approximate the
distribution and composition of bound versus unbound debris by calculating the specific
orbital energy of each particle tracer at the end of the simulations,
then assume bound particles with negative orbital energy subsequently evolve in a collisionless fashion
after having settled onto closed Keplerian orbits.
This allows us to approximate the late-time radial extent of the debris
distribution as $R_{\mathrm{max}}=-GM/(2\epsilon)$, where $R_{\mathrm{max}}$ is effectively the semi-major axis of the 
(closed) orbital trajectories. 

\begin{figure}
\includegraphics[width=0.7\textwidth]{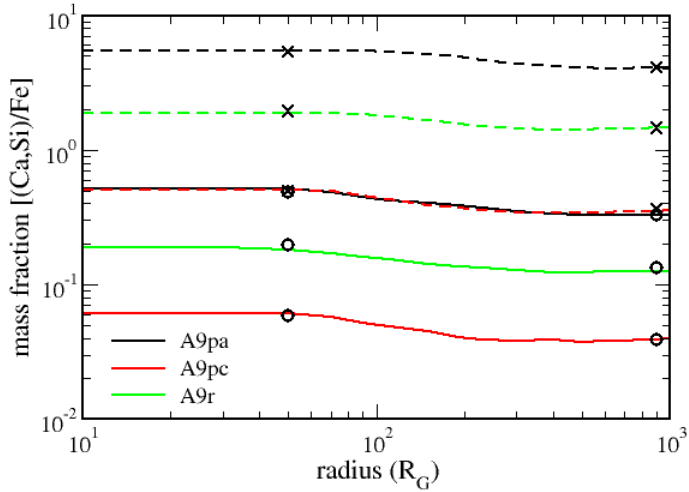}
\caption{
Approximate late-time mass fractions of intermediate mass elements relative
to iron are plotted as a function of radius from the BH in $R_G$ units. Solid (dotted) lines
represent Ca/Fe (Si/Fe) mass fractions. The different colors correspond
to the three different cases A9p-a (black), A9p-c (red), and A9r (green).
Circles (crosses) are  polynomial fits to the Ca/Fe (Si/Fe) fractions at 50 and 900 $R_G$.
}
\label{fig:pp_caoverfe_rad}
\end{figure}

In this way we find the radial distribution of nuclear burn products does not vary much.
For example, the mass of nuclear debris (when catalogued into Fe, Ca and Si groups) 
trapped within $R_{\mathrm{max}} < 10 R_G$ is 
generally only about twice the corresponding group amount traveling on unbound trajectories. As shown in 
Figure \ref{fig:pp_caoverfe_rad}, the ratio of burn products when normalized to total
iron content is just as evenly distributed, varying by less than a factor of two across many orders of
magnitude in radius. For each case we have considered (A9p-a, A9p-c, A9r),
the ratio of unbound intermediate mass elements relative to iron is less than what can be found
in bound debris (by a factor of two), and a clear transition between near and far field distributions
occurs in all calculations between 100 and 200 $R_G$.

The seemingly uniform logarithmic shifts between each of the three cases plotted in Figure \ref{fig:pp_caoverfe_rad}
suggests that perhaps a simple scaling law can be useful for approximating species mass fractions.
For example, applying a least squares fit to a generic polynomial of the form $M_F \propto (\beta^*)^{-n}$, 
the mass ratios can be fit in both the near and far fields rather nicely by 
exponents $n \sim (4.3, ~4.8)$ for the (Ca/Fe, Si/Fe) ratios respectively.
The polynomial fits are performed at two different radii, 50 and 900 $R_G$, 
and represented as symbols in Figure \ref{fig:pp_caoverfe_rad}: circles for Ca/Fe, crosses for Si/Fe.
Of course the range of parameter space used in determining these fits is
rather limited, so we cannot reliably extrapolate this scaling beyond the
intermediate tidal strength interactions ($5 \lesssim \beta^* \lesssim 9$) studied in this report.

\subsection{Isotopic Distribution}
\label{subsec:pf}

Table \ref{tab:pp_pf} presents the final isotopic mass fractions of the dominant stable 
species for both the unbound and total (bound plus unbound) 
stellar matter that survived to the end of the high-resolution runs A9p-a, A9r, and A9p-c. 
The fraction of unbound material in these three runs sum to 43, 45, and 50\% of the total mass respectively.
In the second and third columns we list the element that contained the most mass in the decay chain leading
to the stable species shown (i.e. their radioactive progenitors), and their corresponding mass fractions
in the Sun \citep{Lodders03}.

\begin{deluxetable}{lcccccccc}
\tablecaption{Final Composition Summary \label{tab:pp_pf}}
\tablewidth{0pt}
\tablehead{
\colhead{Nucleus}             & 
\colhead{Progenitor}          & 
\colhead{$X_\odot$}           & 
\colhead{$X_\mathrm{tot}$}    & 
\colhead{$X_\mathrm{tot}$}    &
\colhead{$X_\mathrm{tot}$}    & 
\colhead{$X_\mathrm{unb}$}    & 
\colhead{$X_\mathrm{unb}$}    &
\colhead{$X_\mathrm{unb}$}    \\
                              & 
                              & 
(solar)                       & 
(A9p-a)                       & 
(A9r  )                       &
(A9p-c)                       & 
(A9p-a)                       & 
(A9r  )                       &
(A9p-c)
}
\startdata
$^{4}$He  & He & $2.74\times10^{-1}$ & $1.22\times10^{-2}$ & $1.14\times10^{-2}$ & $3.77\times10^{-2}$ & $8.15\times10^{-3}$ & $5.84\times10^{-3}$ & $3.68\times10^{-2}$ \\
$^{12}$C  & C  & $2.46\times10^{-3}$ & $8.08\times10^{-2}$ & $2.95\times10^{-2}$ & $4.23\times10^{-4}$ & $9.22\times10^{-2}$ & $3.74\times10^{-2}$ & $1.61\times10^{-4}$ \\
$^{16}$O  & O  & $6.60\times10^{-3}$ & $2.02\times10^{-1}$ & $1.40\times10^{-1}$ & $4.85\times10^{-2}$ & $1.78\times10^{-1}$ & $1.21\times10^{-1}$ & $4.59\times10^{-2}$ \\
$^{24}$Mg & Mg & $5.65\times10^{-4}$ & $2.33\times10^{-2}$ & $2.04\times10^{-2}$ & $7.62\times10^{-3}$ & $1.61\times10^{-2}$ & $1.57\times10^{-2}$ & $7.52\times10^{-3}$ \\
\\
$^{28}$Si & Si & $7.55\times10^{-4}$ & $3.06\times10^{-1}$ & $2.81\times10^{-1}$ & $1.54\times10^{-1}$ & $3.06\times10^{-1}$ & $2.70\times10^{-1}$ & $1.39\times10^{-1}$ \\
$^{32}$S  & S  & $3.96\times10^{-4}$ & $1.70\times10^{-1}$ & $1.56\times10^{-1}$ & $8.36\times10^{-2}$ & $1.78\times10^{-1}$ & $1.56\times10^{-1}$ & $7.40\times10^{-2}$ \\
$^{36}$Ar & Ar & $9.13\times10^{-5}$ & $3.99\times10^{-2}$ & $3.83\times10^{-2}$ & $2.00\times10^{-2}$ & $4.19\times10^{-2}$ & $3.78\times10^{-2}$ & $1.73\times10^{-2}$ \\
$^{40}$Ca & Ca & $7.13\times10^{-5}$ & $4.07\times10^{-2}$ & $3.93\times10^{-2}$ & $2.35\times10^{-2}$ & $4.05\times10^{-2}$ & $3.75\times10^{-2}$ & $1.98\times10^{-2}$ \\
\\
$^{44}$Ca & Ti & $1.69\times10^{-6}$ & $1.13\times10^{-3}$ & $7.73\times10^{-4}$ & $9.00\times10^{-4}$ & $4.96\times10^{-4}$ & $3.28\times10^{-4}$ & $6.52\times10^{-4}$ \\
$^{48}$Ti & Cr & $2.51\times10^{-6}$ & $2.71\times10^{-3}$ & $3.83\times10^{-3}$ & $4.42\times10^{-3}$ & $1.20\times10^{-3}$ & $1.62\times10^{-3}$ & $3.81\times10^{-3}$ \\
$^{52}$Cr & Fe & $1.64\times10^{-5}$ & $9.29\times10^{-3}$ & $1.57\times10^{-2}$ & $1.34\times10^{-2}$ & $1.00\times10^{-2}$ & $1.49\times10^{-2}$ & $1.20\times10^{-2}$ \\
$^{56}$Fe & Ni & $1.26\times10^{-3}$ & $9.51\times10^{-2}$ & $2.47\times10^{-1}$ & $5.63\times10^{-1}$ & $1.16\times10^{-1}$ & $2.89\times10^{-1}$ & $5.97\times10^{-1}$ \\
$^{57}$Fe & Ni & $2.96\times10^{-5}$ & $9.79\times10^{-4}$ & $2.55\times10^{-3}$ & $9.28\times10^{-3}$ & $1.04\times10^{-3}$ & $2.67\times10^{-3}$ & $9.74\times10^{-3}$ \\
$^{59}$Co & Cu & $4.01\times10^{-6}$ & $2.22\times10^{-6}$ & $3.76\times10^{-4}$ & $3.26\times10^{-3}$ & $1.66\times10^{-6}$ & $4.25\times10^{-4}$ & $3.40\times10^{-3}$ \\
$^{60}$Ni & Zn & $2.20\times10^{-5}$ & $7.94\times10^{-6}$ & $1.10\times10^{-3}$ & $2.09\times10^{-2}$ & $4.54\times10^{-6}$ & $1.29\times10^{-3}$ & $2.40\times10^{-2}$ \\
$^{64}$Zn & Ge & $1.12\times10^{-6}$ & $9.92\times10^{-9}$ & $2.62\times10^{-5}$ & $9.25\times10^{-4}$ & $6.34\times10^{-13}$& $3.03\times10^{-5}$ & $1.03\times10^{-3}$ \\
\enddata
\end{deluxetable}

All species shown are produced as self conjugate (Z=N) alpha nuclei. Since our initial
compositions are mixtures of alpha particles, essentially $^4$He, $^{12}$C, and $^{16}$O (see Figure 1 in Paper 1),
there is no initial neutron excess, and subsequently minimal production of more neutron rich species.
As the peak temperature increases in ever closer encounters (progressively in the order A9p-a, A9r, Ap9-c) heavier species
are produced with an ever larger fraction of the final mass ending up in $^{56}$Ni.

From the ensemble of tracer particles in runs A9p-a, A9r, and A9p-c we can easily identify and
extract individual thermodynamic trajectories favorable to the production of any specified isotope.
We find, for example, that trajectories winding up with large mass fractions of $^{48}$Cr ($^{56}$Ni) originate from the
outer (inner) portions of the WD, $>5000$ ($<1000$) km from the center core. The initial compositions
of these trajectories consist of roughly 70/30 ratios of
$^4$He/$^{12}$C ($^{16}$O/$^{12}$C) for the $^{48}$Cr ($^{56}$Ni) producing tracks.
Because of the high densities and relatively long burning times, the production of
Ca, Ti, and Cr resembles explosive helium burning \citep{Woosley86} taking place in the outer regions of the WD.
The high densities also allow for efficient conversion of alpha-particles to carbon, resulting in
final alpha-particle mass fractions that were small (less than a few percent) in all scenarios studied here.
Because of the much higher temperatures ($>5\times10^9$ K) experienced by material closer to the center of the WD, the
heavier products (Fe, Co, Ni, Zn) created under these conditions
resemble material that reached nuclear statistical equilibrium at zero neutron excess \citep{Hartmann85}.
$^{64}$Zn (made here as $^{64}$Ge) is at the end of the production chain since proton-unbound species above it prevent further synthesis
in the absence of robust neutrino emission \citep{Pruet06, Frolich06}.

Several species observed in our calculations
are of interest because they contain gamma-ray emiting nuclei in their decay pathways.
The gamma-ray and positron energy from decay of these species, especially $^{56}$Ni and $^{56}$Co, is converted into kinetic energy of the debris or
thermalized with most of the flux appearing at optical wavelengths \cite{Suntzeff02}. While we anticipate that tidal encounters will most likely be first observed
through photometric means, we do point out that with very little to no envelope material surrounding the white dwarf, and the tidal forces acting to produce a spray
rather than an expanding photo-sphere, that the gamma-ray diffision times may be short, thus giving rise to the possibility of
gamma-ray line emission from some of these species that might be observable. We provide below a short discussion of the prominant gamma-lines
and develop approximate bolometric light curves in the next section.

$^{44}$Ti, with a half-life of 60 years, decays through $^{44}$Sc to $^{44}$Ca.
The dominant (99.88\% per $^{44}$Sc decay) gamma-line observed in Type II supernova remnants is the
1157 keV from the 2+ excited state to the 0+ ground state in $^{44}$Ca,
although the 78.36 and 67.86 keV lines from the 0- and 1- states in $^{44}$Sc
have also been observed \citep{Timmes96,Hoffman10}.
Because of the uncertainty in the crucial production and destruction reaction rates affecting its
synthesis (we use the theoretical rates from \cite{rauscher00b}),
our calculated results for $^{44}$Ti synthesis may well be uncertain by a factor of two or more \citep{Hoffman10}.

$^{48}$Cr decays to $^{48}$Ti through $^{48}$V which has a half-life of 16 days.
The dominant gamma lines (983.5 and 1312 keV) observed in this decay chain 
arise via strong E2 transitions in the ground state band of $^{48}$Ti.

$^{52}$Fe decays to $^{52}$Cr through $^{52}$Mn with a 5.6 day half-life. 91.4\% of the time the 6+ ground state of $^{52}$Mn decays to the 6+ excited state in
$^{52}$Cr, and again strong E2 transitions in the ground state band give rise to the most intense gamma lines with energies of 1434, 935.5, and 744.2 keV.

The closest approach scenario (case A9p-c) is an interesting example of an encounter producing copius amounts of the iron group elements
Fe, Co, and Ni along with the rarely seen $^{64}$Zn. Again, each of these species are made as self-conjugate radioactive progenitors.
Prominent gamma-ray lines from the decay of $^{56}$Ni exhibit energies of 0.847 and 1.238,
while those from decay of $^{57}$Ni have energies of 0.122, 0.014, and 0.136 MeV.

$^{59}$Co is produced here as $^{59}$Cu which decays through the long lived (76000 y) $^{59}$Ni, but we would not expect any discernable
gamma-rays (the Cu lifetime is very short with 60\% decaying directly to the ground state of $^{59}$Ni).
The lifetimes of the species that decay to Ni and Zn are also too short to be considered as viable candidates for observation.

Our results suggest a number of radioactive species with potentially observable gamma lines are made in white dwarf tidal disruption
events, and that specific lines and relative line strengths may effectively characterize the encounters.
The collective imaging of $^{56}$Ni, $^{57}$Ni, and possibly $^{44}$Ti, $^{48}$V, $^{52}$Mn (if made
in sufficient quantities) would provide a fairly discriminating diagnostic of interaction strength.

\subsection{Luminosities}
\label{subsec:lum}

\begin{figure}
\includegraphics[width=0.6\textwidth]{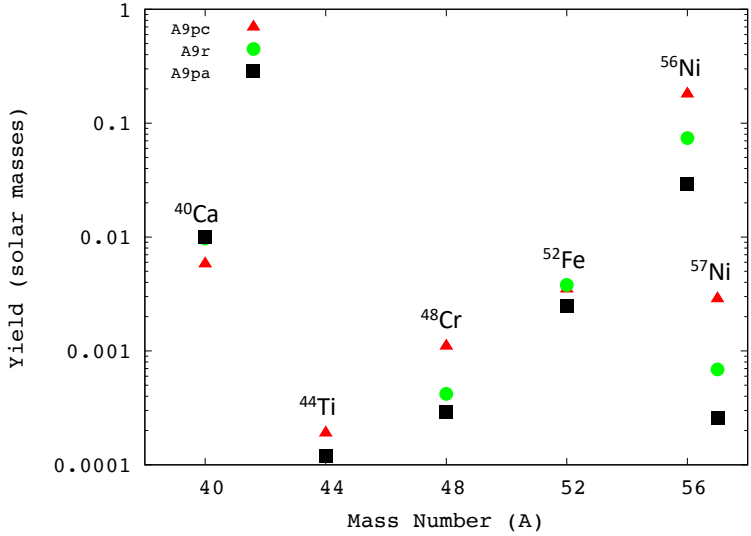}
\caption{
Unbound yields in solar mass units of $^{40}$Ca together with the most abundant radioactive
isotopes produced in cases A9p-c (red triangles), A9r (green circles), and A9-a (black squares).
}
\label{fig:pp_pf}
\end{figure}

In this section we apply simple models to estimate transient luminosities based on unbound $^{56}$Ni yields
calculated by multiplying the mass fractions in Table \ref{tab:pp_pf} with the total
unbound masses in Table \ref{tab:pp_results}. The yields for the three high resolution cases
A9p-c, A9r, and A9p-a are 0.179, 0.078, and 0.030 solar masses respectively.
Figure \ref{fig:pp_pf} plots the $^{56}$Ni yields against some of the other species produced in Table \ref{tab:pp_pf}.

Clearly $^{56}$Ni dominates the yields of the other radioactive species in all three encounter
scenarios, typically by an order of magnitude or more.
In the following we assume that the peak bolometric luminosity is due to energy deposition by the
decay of $^{56}$Ni and $^{56}$Co \citep{Arnett79}, and adopt the procedure described in \citet{Dado15}
to calculate light curves based on four model parameters:
the $^{56}$Ni yield ($M_{56}$), the rise time ($t_r$), a time scale for gamma-ray energy deposition ($t_{\gamma}$), and the
fraction of the energy released into the expanding fireball by positrons from the decay of $^{56}$Co ($A_e$).
These parameters in turn depend on the critical mass for the WD ($M_C$ which we set to 0.6 $M_\odot$),
our calculated $^{56}$Ni yields, and an initial velocity for the fireball expansion ($V_0$).

Table \ref{tab:lcurves} shows the inputs and derived quantites for the three high-resolution runs A9p-c, A9r, and A9p-a.
From these runs we observed a range of expansion velocities ($0.03c \leq V_{0} \leq 0.09c$). The lower limit is consistent with the
assumption that the kinetic energy in the fireball is approximated by the total nuclear energy generated ($V_0=\sqrt{2 e_{nuc}/M_C}$).
We also include the fitted parameters for SN 1992bc \citep[$V_0 ~ 0.05c$, or 15,600 km s$^{-1}$][]{Dado15}.

\begin{deluxetable}{lccccccc}
\tablecaption{Parameters for light curve fits \label{tab:lcurves}}
\tablewidth{0pt}
\tablehead{
\colhead{Object}              &
\colhead{$M$($^{56}$Ni)}      &
\colhead{$V_0$}               &
\colhead{$t_r$}               &
\colhead{$t_\gamma$}          &
\colhead{$t_{peak}$}          &
\colhead{log($L_p$)}          &
\colhead{$\Delta m^{bol}_{15}$}  \\
                              &
($M_\odot$)                   &
($c$)                         &
(d)                           &
(d)                           &
(d)                           &
(erg s$^{-1}$)                &
(mag)
}
\startdata
SN 1992bc   &  ~0.840  & ~0.050 & 12.60 & 26.80 & 17.50 & 43.19 & 0.93 \\
SN 1991bg   &  ~0.110  &      - &     - &     - &     - & 42.32 & 1.42 \\
A9p-c       &   0.179  &  0.039 &  8.03 & 27.31 & 13.50 & 42.68 & 0.88 \\
            &          &  0.060 &  6.50 & 17.91 & 10.52 & 42.73 & 1.32 \\
            &          &  0.090 &  5.31 & 11.94 &  8.52 & 42.75 & 1.84 \\
A9r         &   0.078  &  0.037 &  8.29 & 29.12 & 13.52 & 42.31 & 0.79 \\
            &          &  0.060 &  6.50 & 17.91 & 10.52 & 42.37 & 1.32 \\
A9p-a       &   0.030  &  0.033 &  8.70 & 32.13 & 14.53 & 41.88 & 0.72 \\
            &          &  0.060 &  6.50 & 17.91 & 10.52 & 41.95 & 1.32 \\
\enddata
\end{deluxetable}

\begin{figure}
\includegraphics[width=0.6\textwidth]{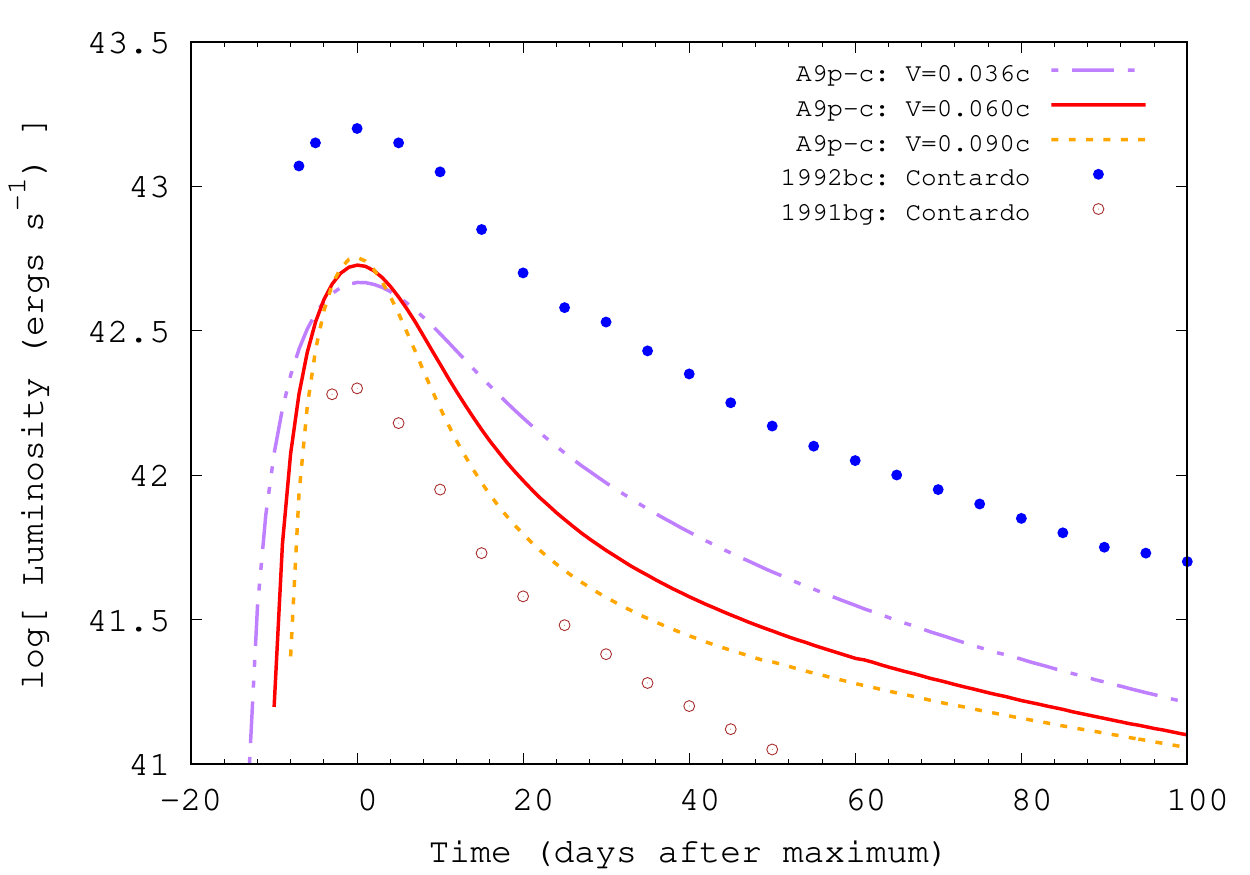} \\
\includegraphics[width=0.6\textwidth]{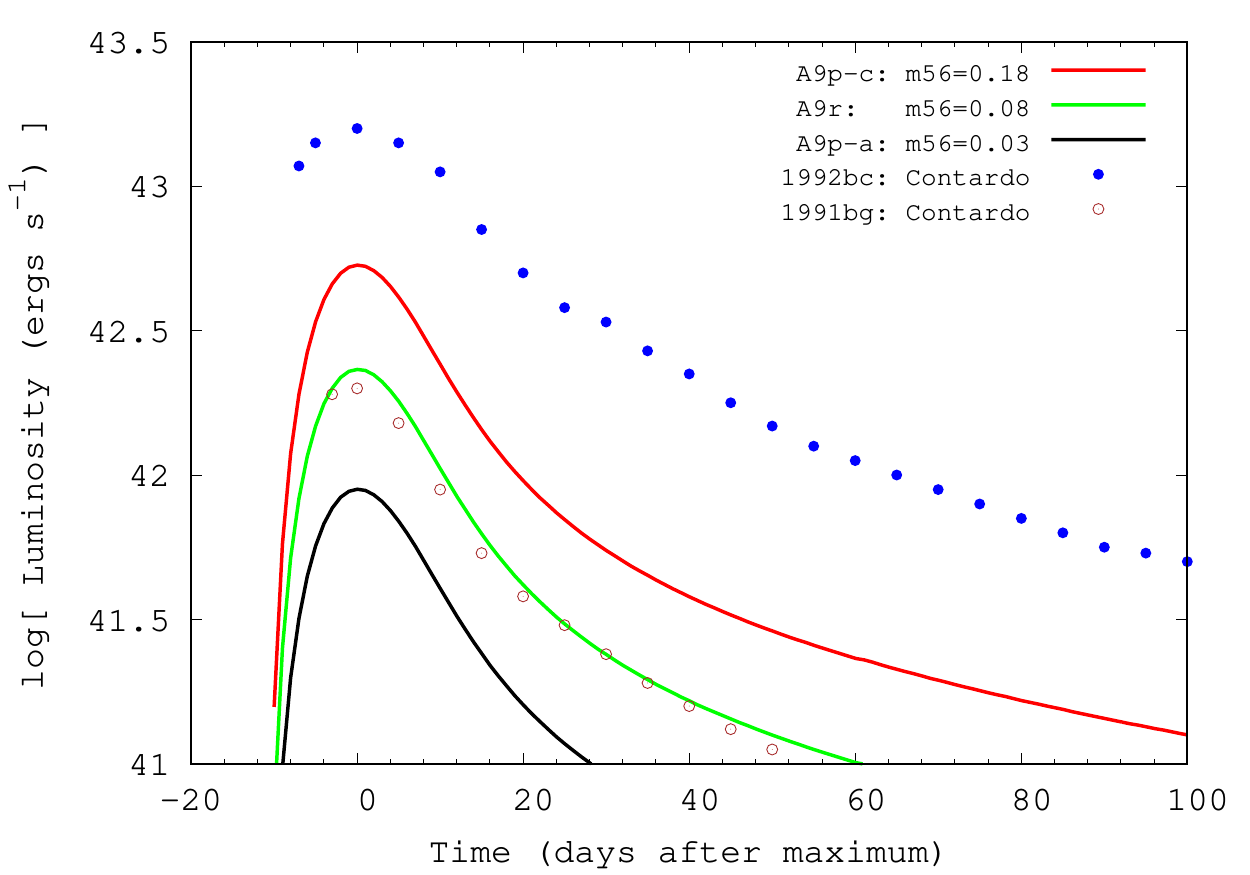} \\
\caption{
Light curves for case A9p-c over a range of expansion velocities $V_0$ (top), and
cases A9p-c, A9r, and A9p-a assuming an average velocity of $0.06c$ (bottom).
Also shown for comparison are the lightcurves from SN 1992bc and SN 1991bg.
}
\label{fig:lcurves}
\end{figure}

In Figure \ref{fig:lcurves} we show our calculated light curves for the three high resolution simulations
together with experimental values of the bolometric
luminosity for the branch-normal SN1992bc and the sub-luminous SN1991bg from \citet{Contardo00}.
The top figure plots a series of light curves for A9p-c assuming a range of matter velocities (0.03c - 0.09c).
This variation mostly affects rise and width of the light curves with a small change in
peak luminosity that favors the higher velocities. The value of $A_e$ also affects the late time behavior, here
we have taken a uniform value of 0.15 for all fits.

The bottom figure shows the light curve fits for Ap9-c, A9r, and A9p-a assuming an averaged
expansion velocity for each of $0.06c$. The differences in peak luminosity
arise from the calculated $^{56}$Ni mass. Interestingly, the $^{56}$Ni mass scales roughly linearly
with interaction parameter $\beta^*$ over the range of encounters considered here so that
$L_b \propto M(^{56}Ni) \propto \beta^*$. Hence we can expect proportionally greater luminosities
from prograde orbits whose characteristically smaller ISCO and MBO constraining radii
allow closer encounters and thus greater interaction strengths.

Figures \ref{fig:lcurves} demonstrate our simulated events are clearly sub-luminous
compared to standard supernovae. The 15 day decline rates
($\Delta m_{15} \approx 2.5 ~\mathrm{log} [L_b(t_p)/L_b(t_p + 15 d)]$) posted in Table \ref{tab:lcurves},
although highly sensitive to the expansion velocity parameter, are generally consistent with standard SN Ia data.
The combination of lower luminosity but comparable decline rates aligns these tidal
scenarios closer to SN 1991bg and SN type Iax events. We stress however that this
analysis is over-simplified and perhaps not entirely appropriate to tidal encounters,
especially considering the anisotropic nature of these events. We defer
more accurate photometric and spectral processing to future work, but note here that luminosities
are expected to be strong functions of viewing angle and may vary by an order
of magnitude or more \citep{Macleod16}.

\section{Conclusions}
\label{sec:conclusions}

We continue to develop and apply a novel numerical methodology 
to the tidal disruption of white dwarf stars from near encounters with intermediate mass black holes.
Here as in Paper 1,
we solve the general relativistic hydrodynamics equations in a Kerr-Schild spacetime to account for relativistic tidal forces,
and utilize a hybrid AMR/moving mesh strategy to move the base grid along Lagrangian fluid lines
while also refining down to a scale of $\sim10^6$ cm on the densest parts of the white dwarf 
as it compresses and responds to tidal forces.
This paper improves on our previous efforts by incorporating a tabular
Helmholtz equation of state with electron degeneracy, significantly upgrading the original idealized 2-component model.
We have also added a particle tracer capability for sampling and tracking stellar material along Lagrangian
fluid lines, modified
the Kerr-Schild metric to include a binding potential of the white dwarf solved dynamically each cycle, and adopted initial composition
profiles from the {\sc MESA} stellar evolution code in place of solving the 
hydrostatic Tolmon-Oppenheimer-Volkof (TOV) equations with a simplified equation of state model. 
As a result of these upgrades, although the primary objective of this paper is to extend previous work to include encounters
with rotating and tilted black holes, we have also taken
this opportunity to revisit a couple of the nonrotating BH encounter scenarios reported in Paper 1.

As a consequence of using {\sc MESA} (not TOV) generated initial stellar models, slight differences
in WD radii and masses implies that the tidal
strength and perihelion radius pairings in this work do not exactly match our previous models.
Differences in the WD masses and equations of state will also affect the efficiency of nuclear reactions. We therefore do not expect
to precisely duplicate our earlier results, but it is interesting and encouraging that they do not differ by much.
For example, projecting our old results to the tidal strength standard $\beta \approx 4.5$, 
we find both of the non-rotating BH encounter scenarios considered in this work agree with the results from Paper 1 to roughly a
factor of two in peak density, peak temperature, nuclear energy release, and element production.

The upgraded particle tracer capability introduced in this work allows us to evaluate the accuracy of the smaller
19-element inline nuclear reaction network by post-processing the traced thermodynamic state
through a much larger 495-isotope network.
The 19- and 495-isotope models agree extremely well: 20\% on total energy release and calcium
production, and better than a factor of two on iron production.

Black hole rotation has very little direct impact on nuclear activity at equivalent tidal strength 
parameterization, not more than 25\% over the tidal strength range considered here, and very
near the margin of numerical uncertainty. When orbital trajectories are adjusted
for relativistic corrections by iterating on the perihelion radius until the WD mass centroids
approach at comparable distances, there is little difference in the total energy release and 
in the final composition of burn products (iron and calcium group elements) even for near maximally rotating BHs
with spin parameter $a=0.9 M_\mathrm{BH}$.
This conclusion holds regardless of black hole mass or orientation of the spin axis relative
to the orbital plane.

Prograde orbital motion has the potential to produce more unbound burn products compared to either
non-rotating BHs or retrograde motion due to smaller ISCO and MBO radii that would allow closer encounters
and thus more efficient nuclear burning and greater luminosity without material plunging immediately into the BH.
As expected, we do find orders of magnitude greater iron production with close prograde
encounters, culminating at about 50 (70)\% mass conversion efficiency in total (unbound) mass elements. 
The mass ratio of unbound intermediate to iron group elements is also affected by prograde alignment,
decreasing by orders of magnitude as increasingly more iron is produced from ever closer encounters
made possible by decreasing marginally bound radii. The Ca/Fe mass ratio for example scales
as $(\beta^*)^{-n}$ with $n\sim 4.3$ in both the near and far field debris over the range
of relativistically corrected interaction strengths $5 \lesssim \beta^* \lesssim 9$.
We additionally find unbound $^{56}$Ni yields, and thus bolometric luminosities, scale
roughly linearly with interaction strength over this same range. The events considered here give
rise to lower peak luminosities ($\sim$0.74 - 4.7 $\times10^{42}$ erg s$^{-1}$) than
standard Ia events, aligning more closely with sub-luminous SN 1991bg and SNe Iax events.

Finally, tidal encounters between WD stars and intermediate black holes have been proposed
as a potential source for observed calcium-rich gap transients \citep{Holcomb13,Sell15}.
Recent observations of the gap transient system candidate SN-2016hnk \citep{Sell18}, however,
has raised doubts to this connection based on a lack of supporting X-ray emission expected from fall-back
accretion immediately following the event.
We nonetheless point out that our previous work did find a strong correlation between
interaction strength and the ratio of intermediate to iron group elements produced in a range of encounters. 
Calcium (iron)-rich debris was a natural outcome from relatively weak (strong) encounters, and
the tidal strength proved to be an effective choking mechanism. 
Although we did not pursue a systematic study of this effect here, the strong dependency of burn product
ratios on tidal strength is evident in these new models as well, particularly in the prograde series of calculations.
One of the motivations for this work was to investigate whether BH spin or spin-orbit alignment 
might provide additional mechanisms for choking nuclear flows and producing incomplete burn environments.
Our findings demonstrate this is not likely, at least for reasonable (less than maximal) rotation rates.

\begin{acknowledgments}

This work was performed in part under the auspices of the U.S. Department of Energy by 
Lawrence Livermore National Laboratory under Contract DE-AC52-07NA27344. It used 
resources from the Extreme Science and Engineering Discovery Environment (XSEDE), 
which is supported by National Science Foundation grant number ACI-1053575. P.C.F.
acknowledges support from National Science Foundation grants AST 1616185 and PHY17-48958. 
\end{acknowledgments}

\software{ Cosmos++ \citep{Anninos05, Anninos12, Anninos17}, MESA \citep{Paxton11}, Torch \citep{Timmes99} }

\listofchanges

\begin{thebibliography}{}
\expandafter\ifx\csname natexlab\endcsname\relax\def\natexlab#1{#1}\fi

\bibitem[{{Abbott} {et~al.}(2016){Abbott}, {Abbott}, {Abbott}, {Abernathy},
  {Acernese}, {Ackley}, {Adams}, {Adams}, {Addesso}, {Adhikari}, \&
  et~al.}]{Abbott16}
{Abbott}, B.~P., {Abbott}, R., {Abbott}, T.~D., {et~al.} 2016, Living Reviews
  in Relativity, 19, 1

\bibitem[{{Anninos} {et~al.}(2018){Anninos}, {Fragile}, {Olivier}, {Hoffman},
  {Mishra}, \& {Camarda}}]{Anninos18}
{Anninos}, P., {Fragile}, P.~C., {Olivier}, S.~S., {et~al.} 2018, \apj, 865, 3

\bibitem[{{Anninos} {et~al.}(2017){Anninos}, {Bryant}, {Fragile}, {Holgado},
  {Lau}, \& {Nemergut}}]{Anninos17}
{Anninos}, P., {Bryant}, C., {Fragile}, P.~C., {et~al.} 2017, \apjs, 231, 17

\bibitem[{{Anninos} {et~al.}(2005){Anninos}, {Fragile}, \& {Salmonson}}]{Anninos05}
{Anninos}, P., {Fragile}, P.~C., \& {Salmonson}, J.~D. 2005, \apj, 635, 723

\bibitem[{{Anninos} {et~al.}(2012){Anninos}, {Fragile}, {Wilson}, \& {Murray}}]{Anninos12}
{Anninos}, P., {Fragile}, P.~C., {Wilson}, J., \& {Murray}, S.~D. 2012, \apj,
  759, 132

\bibitem[{{Arnett} (1979)}]{Arnett79}
{Arnett}, W.~D. 1979, \apjl, 230, L37

\bibitem[{{Brassart} \& {Luminet}(2008)}]{Brassart08}
{Brassart}, M., \& {Luminet}, J.-P. 2008, \aap, 481, 259

\bibitem[{{Camarda} {et~al.}(2009){Camarda}, {Anninos}, {Fragile}, \& {Font}}]{Camarda09}
{Camarda}, K.~D., {Anninos}, P., {Fragile}, P.~C., \& {Font}, J.~A. 2009, \apj, 707, 1610

\bibitem[{{Contardo} {et~al.}(2000){Contardo}, {Leibundgut}, \& {Vacca}}]{Contardo00}
{Contardo}, G., {Leibundgut}, B., \& {Vacca}, W.~D. 2000, \aap, 359, 876

\bibitem[{{Dado} \& {Dar}(2015)}]{Dado15}
{Dado}, S., \& {Dar}, A. 2015, \apj, 809, 32

\bibitem[{{Dong} {et~al.}(2007){Dong}, {Wang}, {Yuan}, {Shan}, {Zhou}, {Fan},
  {Dou}, {Wang}, {Wang}, \& {Lu}}]{Dong07}
{Dong}, X., {Wang}, T., {Yuan}, W., {et~al.} 2007, \apj, 657, 700

\bibitem[{{Evans} {et~al.}(2015){Evans}, {Laguna} \& {Eracleous}}]{Evans15}
{Evans}, C., {Laguna}, P., \& {Eracleous}, M. 2015, \apj, 805, L19

\bibitem[{{Farr} {et~al.}(2017){Farr}, {Stevenson}, {Miller}, {Mandel}, {Farr} \& {Vecchio}}]{Farr17}
{Farr}, W.~M., {Stevenson}, S., {Miller}, M.~C., {Mandel}, I., {Farr}, B. \& {Vecchio}, A.
  2017, \nat, 548, 426

\bibitem[{{Fragile} {et~al.}(2012){Fragile}, {Gillespie}, {Monahan},
  {Rodriguez}, \& {Anninos}}]{Fragile12}
{Fragile}, P.~C., {Gillespie}, A., {Monahan}, T., {Rodriguez}, M., \&
  {Anninos}, P. 2012, \apjs, 201, 9

\bibitem[{{Fragile} {et~al.}(2014){Fragile}, {Olejar}, \&
  {Anninos}}]{Fragile14}
{Fragile}, P.~C., {Olejar}, A., \& {Anninos}, P. 2014, \apj, 796, 22

\bibitem[Fr\"ohlich  et. al.~(2006)]{Frolich06}
Fr\"ohlich, C., Mart\'{\i}nez-Pinedo, G., Liebend\"orfer, M.,
Thielemann, F.-K., Bravo, E., Hix, W.~R., Langanke, K.-H., \& Zinner, N.~T. \ 2006,
Phys. Rev. Lett. 96, 142502

\bibitem[{{Gafton} \& {Rosswog}(2019){Gafton} \& {Rosswog}}]{Gafton19}
{Gafton}, E. \& {Rosswog}, S. 2019, ArXiv e-prints, arXiv:1903.09147

\bibitem[{{Gammie} {et~al.}(2004){Gammie}, {Shapiro} \& {McKinney}}]{Gammie04}
{Gammie}, C.~F., {Shapiro}, S.~L., \& {McKinney}, J.~C. 2004, \apj, 602, 312

\bibitem[{{Gebhardt} {et~al.}(2002){Gebhardt}, {Rich}, \& {Ho}}]{Gebhardt02}
{Gebhardt}, K., {Rich}, R.~M., \& {Ho}, L.~C. 2002, \apjl, 578, L41

\bibitem[{{Gebhardt} {et~al.}(2005){Gebhardt}, {Rich}, \& {Ho}}]{Gebhardt05}
---. 2005, \apj, 634, 1093

\bibitem[{{Gerssen} {et~al.}(2002){Gerssen}, {van der Marel}, {Gebhardt},
{Guhathakurta}, {Peterson}, \& {Pryor}}]{Gerssen02}
{Gerssen}, J., {van der Marel}, R.~P., {Gebhardt}, K., {et~al.} 2002, \aj, 124, 3270

\bibitem[{{Gerssen} {et~al.}(2003){Gerssen}, {van der Marel}, {Gebhardt},
{Guhathakurta}, {Peterson}, \& {Pryor}}]{Gerssen03} ---. 2003, \aj, 125, 376

\bibitem[{{Haas} {et~al.}(2012){Haas}, {Shcherbakov}, {Bode}, \& {Laguna}}]{Haas12}
{Haas}, R., {Shcherbakov}, R.~V., {Bode}, T., \& {Laguna}, P. 2012, \apj, 749, 117

\bibitem[{{Hartmann} {et~al.}(1985){Hartmann}, {Woosley}, \& {El Eid}}]{Hartmann85}
{Hartmann}, D., {Woosley}, S.~E., \& {El Eid}, M. 1985, \apj, 297, 837

\bibitem[{{Hoffman} {et~al.}(2010){Hoffman et al.}}]{Hoffman10}
{Hoffman}, R.~D. \& {et al.} 2010, \apj, 715, 1383

\bibitem[{{Holcomb} {et~al.}(2013){Holcomb}, {Guillochon}, {De Colle}, \&
  {Ramirez-Ruiz}}]{Holcomb13}
{Holcomb}, C., {Guillochon}, J., {De Colle}, F., \& {Ramirez-Ruiz}, E. 2013, \apj, 771, 14

\bibitem[{{Kawana} {et~al.}(2018){Kawana}, {Tanikawa}, \& {Yoshida}}]{Kawana18}
{Kawana}, K., {Tanikawa}, A., \& {Yoshida}, N. 2017, \mnras 477, 3449

\bibitem[{{Lodders}(2003){Lodders}}]{Lodders03}
{Lodders}, K. 2003, \apj, 591, 1220

\bibitem[{{Luminet} \& {Carter}(1986)}]{Luminet86}
{Luminet}, J.-P., \& {Carter}, B. 1986, \apjs, 61, 219

\bibitem[{{Luminet} \& {Pichon}(1989)}]{Luminet89}
{Luminet}, J.-P., \& {Pichon}, B. 1989, \aap, 209, 103

\bibitem[{{MacLeod} {et~al.}(2016){MacLeod}, {Guillochon}, {Ramirez-Ruiz}, {Kasen}, \& {Rosswog}}]{Macleod16}
{MacLeod}, M., {Guillochon}, J., {Ramirez-Ruiz}, E., {Kasen}, D., \& {Rosswog}, S. 2016, \apj, 819, 3

\bibitem[{{McClintock} {et~al.}(2014){McClintock}, {Narayan} \& {Steiner}}]{McClintock14}
{McClintock}, J.~E., {Narayan}, R., \& {Steiner}, J.~F. 2014, \ssr, 183, 295

\bibitem[{{Paxton} {et~al.}(2011){Paxton}, {Bildsten}, {Dotter}, {Herwig},
  {Lesaffre}, \& {Timmes}}]{Paxton11}
{Paxton}, B., {Bildsten}, L., {Dotter}, A., {et~al.} 2011, \apjs, 192, 3

\bibitem[{{Pruet} {et~al.}(2006){Pruet et al.}}]{Pruet06}
{Pruet}, J. \& {et al.} 2006, \apj, 644, 1028

\bibitem[{Rauscher \& Thielemann(2000)}]{rauscher00b}
Rauscher, T. \& Thielemann, F. 2001, At. Data Nucl. Data Tables, 75, 1

\bibitem[{{Rees}(1988)}]{Rees88}
{Rees}, M.~J. 1988, \nat, 333, 523

\bibitem[{{Reynolds}(2014)}]{Reynolds14}
{Reynolds}, C.-S. 2014, \ssr, 183, 277

\bibitem[{{Rosswog} {et~al.}(2009){Rosswog}, {Ramirez-Ruiz}, \&
  {Hix}}]{Rosswog09}
{Rosswog}, S., {Ramirez-Ruiz}, E., \& {Hix}, W.~R. 2009, \apj, 695, 404

\bibitem[{{Seitenzahl} {et~al.}(2009){Seitenzahl}, {Meakin}, {Townsley}, {Lamb} \& {Truran} }]{Seitenzahl09}
{Seitenzahl}, I.~R., {Meakin}, C.~A., {Townsley}, D.~M., {Lamb}, D.~Q., \& {Truran}, J.~W. 2009, \apj, 696, 515

\bibitem[{{Sell} {et~al.}(2015){Sell}, {Maccarone}, {Kotak}, {Knigge}, \& {Sand}}]{Sell15}
{Sell}, P.~H., {Maccarone}, T.~J., {Kotak}, R., {Knigge}, C., \& {Sand}, D.~J. 2015, \mnras, 450, 4198

\bibitem[{{Sell} {et~al.}(2018){Sell}, {Arur}, {Maccarone}, {Kotak}, {Knigge}, {Sand} \& {Valenti}}]{Sell18}
{Sell}, P.~H., {Arur}, K., {Maccarone}, T.~J., {Kotak}, R., {Knigge}, C., {Sand}, D.~J., \& {Valenti}, S. 2018, \mnras, 475, L111

\bibitem[{{Stone} {et~al.}(2013){Stone}, {Sari}, \& {Loeb}}]{Stone13}
{Stone}, N., {Sari}, R., \& {Loeb}, A. 2013, \mnras, 435, 1809

\bibitem[{{Suntzeff}(2002)}]{Suntzeff02}
{Suntzeff}, N.~B. 2002, in From Twilight to Highlight, The Physics of Supernovae, ed. W. Hillebrandt, B. Leibundgut, 183

\bibitem[{{Tanikawa} {et~al.}(2017){Tanikawa}, {Sato}, {Nomoto}, {Maeda},
  {Nakasato}, \& {Hachisu}}]{Tanikawa17}
{Tanikawa}, A., {Sato}, Y., {Nomoto}, K., {et~al.} 2017, \apj, 839, 81

\bibitem[{{Thorne}(1974)}]{Thorne74}
{Thorne}, K.-S. 1974, \apj, 191, 507

\bibitem[{{Timmes} {et~al.}(1996){Timmes}, {Woosley}, {Hartmann}, \& {Hoffman}}]{Timmes96}
{Timmes}, F.~X., {Woosley}, S.~E., {Hartmann}, D.~H., \& {Hoffman}, R.~D. 1996, \apj, 464, 332

\bibitem[{{Timmes}(1999)}]{Timmes99}
{Timmes}, F.~X. 1999, \apjs, 124, 241

\bibitem[{{Timmes} {et~al.}(2000a){Timmes}, \& {Swesty}}]{Timmes00a}
{Timmes}, F.~X., \& {Swesty}, F.~D. 2000, \apjs, 126, 501

\bibitem[{{Timmes} {et~al.}(2000b){Timmes}, {Hoffman}, \& {Woosley}}]{Timmes00b}
{Timmes}, F.~X., {Hoffman}, R.~D., \& {Woosley}, S.~E. 2000, \apjs, 129, 377

\bibitem[{{Weaver} {et~al.}(1978){Weaver}, {Zimmerman}, \& {Woosley}}]{Weaver78}
{Weaver}, T.~A., {Zimmerman}, G.~B., \& {Woosley}, S.~E. 1978, \apj, 225, 1021

\bibitem[{{Woosley}(1986){Woosley}}]{Woosley86}
{Woosley}, S.~E. 1986, in Nucleosynthesis and Chemical Evolution,
ed. B. Hauck, A. Maeder, G. Meynet, 74

\end{thebibliography}
\end{document}